\documentclass[acmlarge,screen]{acmart}
\AtBeginDocument{%
  }

\setcopyright{cc}
\setcctype{by}
\acmJournal{IMWUT}
\acmYear{2026} \acmVolume{10} \acmNumber{3} \acmArticle{81}
\acmMonth{9} \acmDOI{10.1145/3831952}

\usepackage{enumitem}
\usepackage{tabularx}
\usepackage{booktabs}
\usepackage{multirow}
\usepackage{adjustbox}
\usepackage{tabularx}
\usepackage{listings}

\newif\ifcomment
\commentfalse

\ifcomment
\newcommand{\hide}[1]{}
\newcommand{\note}[1]{\textcolor{blue}{<< #1 >>}}
\newcommand{\added}[1]{\textcolor[rgb]{0.30,0.55,0.85}{#1}}

\newcommand{\cut}[1]{\textcolor[rgb]{0.5,0.5,0.5}{CUT: #1}}

\newcommand{\todo}[1]{\textcolor{red}{TODO: #1}}

\newcommand{\deleted}[1]{}

\newcommand{\runze}[1]{\textcolor[rgb]{0.5,0.2,0.2}{RUNZE: #1}}
\newcommand{\zsd}[1]{\textcolor[rgb]{0.5,0.1,0.8}{ZSD: #1}}
\newcommand{\lp}[1]{\textcolor[rgb]{0.6,0.6,1.0}{Linping: #1}}
\newcommand{\jussi}[1]{\textcolor[rgb]{0.2,0.6,1.0}{Jussi: #1}}

\else
\newcommand{\hide}[1]{}
\newcommand{\note}[1]{}
\newcommand{\added}[1]{#1}

\newcommand{\cut}[1]{}

\newcommand{\todo}[1]{}

\newcommand{\deleted}[1]{}

\newcommand{\runze}[1]{}
\newcommand{\zsd}[1]{}
\newcommand{\lp}[1]{}
\newcommand{\jussi}[1]{}

\fi

\newcommand{\meansd}[2]{$M = #1,\: SD = #2$}

\renewcommand{\quote}[1]{``\textit{#1}''}
\newcommand{\quoteby}[2]{``\textit{#2} (#1)''}

\newcommand{\wearableAI}[0]{\textit{Wearable Creative AI}}

\newcommand{\enjoyment}[0]{\textit{Enjoyment}}
\newcommand{\lowDistraction}[0]{\textit{Low Distraction}}

\newcommand{\taskDuration}[0]{\textit{Task Duration}}

\newcommand{\control}[0]{\textit{Control}}
\newcommand{\unexpectedness}[0]{\textit{Unexpectedness}}
\newcommand{\unnoticedDetail}[0]{\textit{Unnoticed Detail}}
\newcommand{\useOfDailyObservation}[0]{\textit{Use of Daily Observation}}

\newcommand{\deepenEnvironment}[0]{\textit{Deepening Connection with Environment}}
\newcommand{\exploration}[0]{\textit{Creative Exploration}}

\newcommand{\outputQuality}[0]{\textit{Output Quality}}
\newcommand{\worthEffort}[0]{\textit{Worth Effort}}
\newcommand{\expressiveness}[0]{\textit{Expressiveness}}

\definecolor{mypurple}{RGB}{128, 0, 255} %

\newcommand{\DPone}{\protect\hyperref[sec:study1:design_principle]{\textbf{\textit{DG1}}}}
\newcommand{\DPtwo}{\protect\hyperref[sec:study1:design_principle]{\textbf{\textit{DG2}}}}
\newcommand{\DPthree}{\protect\hyperref[sec:study1:design_principle]{\textbf{\textit{DG3}}}}

\newcommand{\CRAFT}[0]{\textit{CRAFT}}

\begin{document}

\title[\CRAFT{}]{\CRAFT{}: Exploring \wearableAI{} on Smart Glasses for Fiction Writing in Real-World Contexts}

\author{Runze Cai}
\orcid{0000-0003-0974-3751}
\affiliation{%
  \institution{National University of Singapore}
  \department{Synteraction Lab, School of Computing}
  \city{Singapore}
  \country{Singapore}
}
\email{runze.cai@u.nus.edu}

\author{Yuxuan Huang}
\authornotemark[2]
\orcid{0000-0002-9452-1506}
\affiliation{%
  \institution{City University of Hong Kong}
  \city{Hong Kong}
  \country{China}
}
\email{yhuang573-c@my.cityu.edu.hk}

\author{Lin-Ping Yuan}
\authornotemark[2]
\orcid{0000-0001-6268-1583}
\affiliation{%
  \institution{The Hong Kong University of Science and Technology}
  \department{Department of Computer Science and Engineering}
  \city{Hong Kong}
  \country{China}
}
\email{yuanlp@cse.ust.hk}

\author{Kexin Xiang}
\orcid{0009-0009-7188-103X}
\affiliation{%
  \institution{City University of Hong Kong}
  \city{Hong Kong}
  \country{China}
}
\email{kxiang9-c@my.cityu.edu.hk}

\author{David Hsu}
\orcid{0000-0002-2309-4535}
\affiliation{%
  \institution{National University of Singapore}
  \department{School of Computing, Smart Systems Institute}
  \city{Singapore}
  \country{Singapore}
}
\email{dyhsu@comp.nus.edu.sg}

\author{Collier Nogues}
\orcid{0000-0002-2321-2028}
\affiliation{%
  \institution{The Chinese University of Hong Kong}
  \department{Department of English}
  \city{Hong Kong}
  \country{China}
}
\email{cnogues@cuhk.edu.hk}

\author{Jussi Holopainen}
\orcid{0000-0001-6264-0407}
\affiliation{%
  \institution{City University of Hong Kong}
  \city{Hong Kong}
  \country{China}
}
\email{jholopai@cityu.edu.hk}

\author{Shengdong Zhao}
\orcid{0000-0001-7971-3107}
\authornotemark[1]
\affiliation{%
  \institution{City University of Hong Kong}
  \department{School of Creative Media \& Department of Computer Science}
  \city{Hong Kong}
  \country{China}
}
\email{shengdong.zhao@cityu.edu.hk}

\makeatletter
\g@addto@macro\@authornotes{%
  \footnotetext[2]{Co-author equal contribution.}%
  \footnotetext[1]{Corresponding author.}%

}
\makeatother

\authorsaddresses{%
Authors' Contact Information:
\href{https://orcid.org/0000-0003-0974-3751}{Runze Cai},
National University of Singapore,
Synteraction Lab, School of Computing,
Singapore, Singapore,
runze.cai@u.nus.edu;
\href{https://orcid.org/0000-0002-9452-1506}{Yuxuan Huang},
City University of Hong Kong,
Hong Kong, China,
yhuang573-c@my.cityu.edu.hk;
\href{https://orcid.org/0000-0001-6268-1583}{Lin-Ping Yuan},
The Hong Kong University of Science and Technology,
Department of Computer Science and Engineering,
Hong Kong, China,
yuanlp@cse.ust.hk;
\href{https://orcid.org/0009-0009-7188-103X}{Kexin Xiang},
City University of Hong Kong,
Hong Kong, China,
kxiang9-c@my.cityu.edu.hk;
\href{https://orcid.org/0000-0002-2309-4535}{David Hsu},
National University of Singapore,
School of Computing, Smart Systems Institute,
Singapore, Singapore,
dyhsu@comp.nus.edu.sg;
\href{https://orcid.org/0000-0002-2321-2028}{Collier Nogues},
The Chinese University of Hong Kong,
Department of English,
Hong Kong, China,
cnogues@cuhk.edu.hk;
\href{https://orcid.org/0000-0001-6264-0407}{Jussi Holopainen},
City University of Hong Kong,
Hong Kong, China,
jholopai@cityu.edu.hk;
\href{https://orcid.org/0000-0001-7971-3107}{Shengdong Zhao}
(corresponding author),
City University of Hong Kong,
School of Creative Media \& Department of Computer Science,
Hong Kong, China,
shengdong.zhao@cityu.edu.hk.
\newline*For correspondence, please contact Jussi Holopainen.
}
\renewcommand{\shortauthors}{Cai et al.}

\begin{abstract}
Creative writing increasingly integrates AI assistance, yet current tools miss in-situ moments when writers draw inspiration from real-world experiences. We envision Context-aware Reality–Fiction Transformation (\CRAFT{}), an approach for AI glasses that translates daily experiences into fiction narratives. 
We explored its desirability, feasibility, and \added{potential} viability through three studies. Interviews with nine writers yielded desires and three design goals: 1) augmenting in-situ perception to bridge reality–fiction gaps, 2) promoting authenticity grounded in real-world experiences while maintaining fictionalization, and 3) preserving creative agency, enjoyment, and life-art boundaries. Co-design workshops with 16 writers and researchers operationalized these goals into concrete interaction mechanisms using a technology probe. We then conducted supported field trials with eight writers across 24 sessions using a refined probe, revealing writer-perceived benefits (e.g., enriched fictional ideas from serendipitous encounters), emergent practices (e.g., micro-creation), and \added{design considerations for future} sustained use. We contribute design explorations for the \CRAFT{} approach, offering design implications and empirical insights on ubiquitous human–AI creative collaboration in everyday life.
\end{abstract}




\maketitle

\section{INTRODUCTION}
\label{sec:introduction}
\begin{figure*}[tbph]
    \centering
  \includegraphics[width=0.73\textwidth]{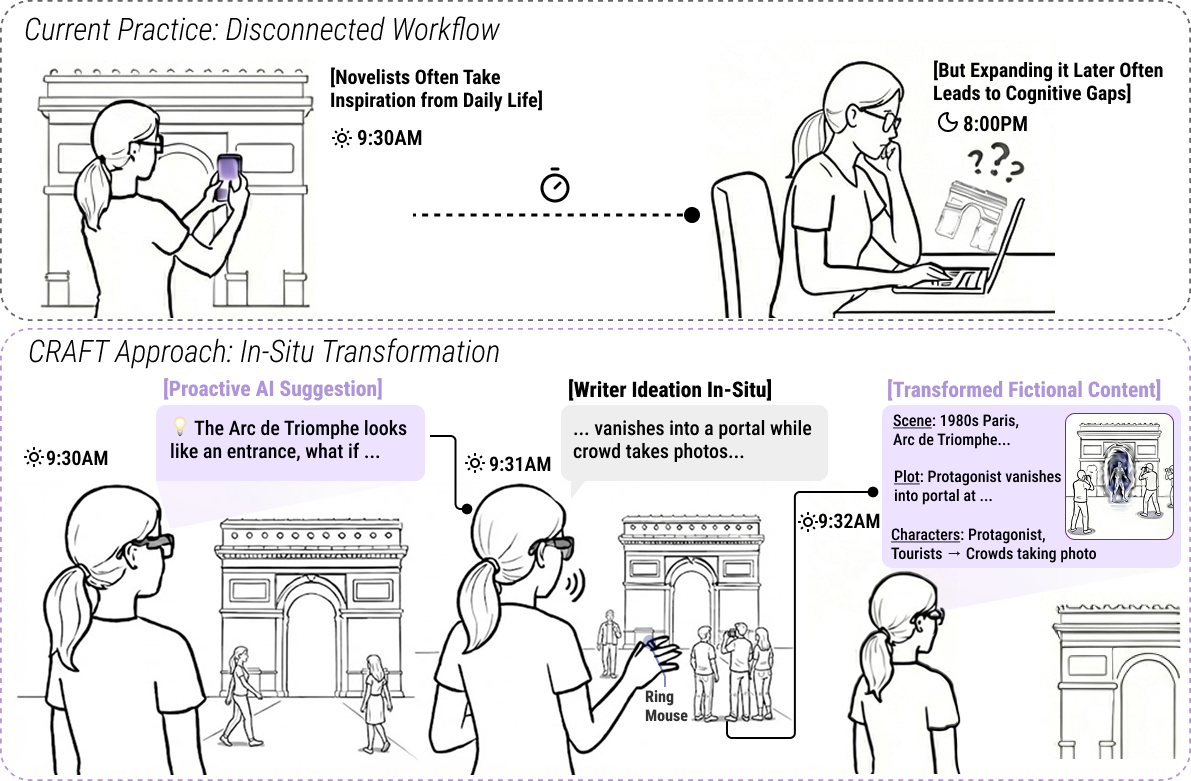}
  \caption{
  \textbf{Top:} Novelists often draw inspiration from daily life (e.g., observing a landmark) but face cognitive gaps (e.g., memory fade of immersive details that a phone recording cannot fully capture) when trying to expand these inspiration moments into writing later.
  \textbf{Bottom:} \CRAFT{} aims to enable a more seamless in-situ transformation of observation into fiction through a \wearableAI{} on smart glasses.
  (1) By proactively analyzing the writer's environment and writing plans, \CRAFT{} suggests a glanceable creative spark ("The Arc looks like an entrance"), filling a plot gap in her fantasy novel about how a protagonist travels between the normal and magical worlds.
  (2) Inspired, the writer clicks her ring mouse and verbally ideates on the spot.
  (3) \CRAFT{} transforms the realistic scene and narration into fictional content with an image, displayed on her smart glasses to maintain immersion in both the physical and fictional worlds for her continuous story development.}
  \Description{A two-part diagram comparing traditional writing workflows with the CRAFT system. The top half shows a writer walking, taking a photo of an arch with a phone, and then struggling to write about it later on a laptop with question marks. The bottom half shows the same writer wearing smart glasses. First, the glasses show a proactive AI suggestion: 'The Arc de Triomphe looks like an entrance, what if...'. Second, the writer speaks '...vanishes into a portal while the crowd takes photos...'. Third, the glasses display 'Transformed Fictional Content', including a scene description of 1980s Paris, a plot about a portal, and character details.}
  \label{fig:teaser}
\end{figure*}

\noindent{}Fiction writers often derive ideas from everyday life—drawing authenticity and emotional resonance from observing and sensing their surroundings, and transforming lived moments into characters, settings, and plots~\cite{williams_writer_2013, doleel_truth_1980, saxton_true_2020, bruner1986actual, hannula_fictionalising_2003}. Yet writers find it difficult to write in-situ when inspiration strikes due to the cognitive demands of creative transformation \cite{paton_dynamic_2016,becker2006review, lee_design_2024, gero2022design} and environmental and technological limitations~\cite{cai_pandalens_2024}. As a compromise, experienced writers take notes, photos, or voice memos to capture inspirational moments~\cite{ahrens2022take, zhao_making_2025}, but the detailed atmosphere and contextual richness of immersive experiences are often lost when recalled later~\cite{zhao_making_2025} (Figure~\ref{fig:teaser}-Top). Consequently, fiction writing remains disconnected from the inspiring moments and contexts, confined largely to stationary, dedicated environments \cite{lee_relationships_2023}.

Recent advances in heads-up, wearable AI assistants on AR glasses~\cite{zhao2023headsup, cai_pandalens_2024, cai_aiget_2025, lee2024gazepointar, wang2024G-VOILA, zhu2025agentar} have demonstrated the potential to enable ubiquitous writing with real-time support. These systems can see what users see and communicate intelligently through voice, successfully supporting task-focused writing (e.g., emails~\cite{zhou2024glassmail}) and reality-based documentation (e.g., travel blogs~\cite{cai_pandalens_2024}) on the go. However, extending these successes to fiction writing is not straightforward, as it presents unique challenges~\cite{cohn2000distinction}. Fiction is more open-ended and typically much longer, and it requires creatively blending reality with imagination—a cognitive transformation process that is more complex than composing structured messages or transcribing daily observations verbatim.

In this work, we explore smart glasses—as an emerging wearable AI form factor—for fiction writing and envision \emph{Context-aware Reality–Fiction Transformation} (\CRAFT{}), an approach that supports translating the writer's lived experience and environmental observation into fictional material \emph{in-situ} with AR smart glasses and AI assistance (referred to as \wearableAI{}). 
Figure~\ref{fig:teaser} (bottom) illustrates one usage scenario: while observing the Arc de Triomphe, a novelist receives a proactive suggestion connecting the arc's shape to a plot gap in her fantasy novel. She then speaks her ideas, and \wearableAI{} transforms her observations into fictional content, supported by an overlay that partially blends the real scene with fictional elements. By further querying historical details about 1980s Paris, she develops a plot in which her protagonist travels between the present and the 1980s through the arc.
This observe$\rightarrow$suggest$\rightarrow$express$\rightarrow$transform$\rightarrow$re-observe loop maintains immersion in both physical and fictional worlds while capturing contextual richness.

To systematically explore the \CRAFT{} approach, we follow the \emph{Design Thinking} methodology~\cite{brown2008design, mesban2023hybrideval}, which evaluates desirability (\emph{RQ1: what users want}), feasibility (\emph{RQ2: how to build}), and viability (\emph{RQ3: real-world benefits and long-term sustainability})\footnote{In the original definition from IDEO~\cite{ideo_prototype_new_business}, viability focuses on business aspects; in our context, we adapt it to ecological/real-world usage benefits following prior HCI research on viability \cite{haag2025last, schuller2024generating}.}. Our studies therefore prioritize the exploration of design and usage opportunities over comparative evaluation against alternative technologies. 
We conduct three complementary studies that progressively address each dimension (Figure~\ref{fig:study_overview}): \textbf{Study 1} interviews 9 experienced writers to understand current practices and desired capabilities, confirming their desirability and yielding three design goals that informed a technology probe to explore design possibilities~\cite{hutchinson2003technology}. \textbf{Study 2} uses co-design workshops with 16 participants to develop concrete interaction designs using the probe, addressing feasibility through \emph{similarity-based prioritization}, \emph{multi-layered perceptual augmentation}, and mechanisms for enjoyment and boundary management. \textbf{Study 3} deploys the refined probe with 8 writers across 24 supported field trials, producing 8 stories, exploring potential viability through author-perceived usage benefits, practice, and long-term design considerations.

Building on these studies, we contribute design explorations of the \CRAFT{} approach for future development of embodied human–AI creative collaboration in ubiquitous environments, with conceptual and empirical contributions:

(1) We identify how writers transform reality into fiction and articulate three design goals: 1) \textit{augmenting in-situ perception} to bridge cognitive gaps between reality and fiction; 2) \textit{promoting the authenticity triad} (factual, logical \& behavioral, and emotional \& psychological authenticity~\cite{doleel_truth_1980, saxton_true_2020, bruner1986actual, hannula_fictionalising_2003}) by utilizing real-world context while maintaining a degree of fictionalization; and 3) \textit{preserving creative agency, enjoyment, and boundaries}~\cite{cherry_csi_2014, holopainen2025infinity, kim2024authors} in daily life.

(2) We suggest design implications for in-situ fiction creation, progressing from general requirements derived from interviews, to specific interaction mechanisms, and long-term and ecological considerations through co-design workshops and supported field trials.

(3) We demonstrate author-perceived benefits and creative practices enabled by \wearableAI{} through continued usage in the real world, such as enriched and authentic fiction ideas from serendipitous daily encounters, micro-creation through brief moments to overcome time barriers, and embodied character perspectives grounded in physical context.

\section{BACKGROUND AND RELATED WORK}

We seek to understand: (1) how fiction writers currently draw from lived experience, (2) the trade-offs of existing AI tools for fiction writing, and (3) what current wearable systems enable and what remains unexplored. Accordingly, we review fiction writing as embodied practice (sec~\ref{sec:rw_embodied}), AI-assisted creative writing tools (sec~\ref{sec:rw_ai_systems}), and mobile/wearable/heads-up computing (sec~\ref{sec:rw_wearable}) in the following sections.

\subsection{Creative and Fiction Writing as an Embodied and Situated Practice}
\label{sec:rw_embodied}

Creative and fiction writing is more than a technical skill; it fulfills central psychological and social functions by enabling people to explore emotions, identities, and collective imagination \cite{dutton_pleasures_2004}. 
Prior research has framed writing from multiple perspectives, including cognitive-process, social-cognitive, and sociocultural theories~\cite{leggette2015review}. 
\added{Within the cognitive-process tradition, we use the Flower \& Hayes model as a design lens for its account of writing as recursive planning, translating, and reviewing~\cite{paton_dynamic_2016, lee_design_2024, gero2022design, becker2006review, flower1981cognitive}. 
This lens helps characterize how \CRAFT{} supports writers in noticing environmental cues, planning how to use them, translating observations into fictional material, and reviewing how that material fits within an evolving story world. 
It also helps explain the cognitive load of in-situ fiction writing: writers must coordinate perception, memory, imagination, narrative consistency, and self-evaluation while engaging with the environment.  
This load often leads writers to seek private or quiet spaces to minimize distraction and enhance creative focus and idea generation~\cite{lee_relationships_2023}.}

However, creativity is not confined to the desk. Exposure to natural environments and embodied practices such as walking can stimulate associative thinking and divergent ideation \cite{yeh2022influence, jabr2014walking}. Creative writing scholarship similarly emphasizes this embodied dimension: writers act as ``hunters and gatherers in the real world'' \cite{krauth2006domains}, using daily walks for ``perceptual decoupling'' that sparks new narrative possibilities \cite{williams_writer_2013}. Ingold’s dwelling perspective further situates imagination in attentive engagement with landscape and sensory impressions \cite{ingold2010ways, ingold2021perception}.

These accounts demonstrate that creative writing is both cognitively demanding and deeply embedded in embodied, situated interaction with the world, highlighting a need for tools that more closely align with the two aspects.

\subsection{AI-Assisted Creative Writing}
\label{sec:rw_ai_systems}
To reduce writers' cognitive load and boost creativity, emerging AI, especially large language models (LLMs), is utilized to support creative writing, covering various genres including lyrics \cite{kim_amuse_2025}, screenplays \cite{mirowski2023co}, poetry \cite{huang2025lyric}, creative science writing \cite{gero2022sparks, zhang2025revtogether}, stories and fiction \cite{yuan2022wordcraft, yu2024lfed, qin2024charactermeet}. Commercial tools (e.g., Sudowrite \cite{fang2024sudowrite}) and research systems \cite{qin2024charactermeet, mirowski2023co, carrera2025nabokov, fan2024storyprompt} offer features from inline generation to story outlining, character construction, and world-building \cite{guo_pen_2024, wan2024felt}. These tools can enhance the creativity of individual ideas and synthesize diverse concepts into coherent responses \cite{lee_empirical_2024}, with LLM-authored texts exhibiting rich perceptual detail and thematic complexity that are sometimes comparable to human writing \cite{orwig2024language, elias_rethinking_2025}.

However, current LLMs can also produce homogeneous patterns or plots~\cite{kim2024authors, beguvs2024experimental}, struggle to capture psychological depth or artistic qualities valued in literature~\cite{elias_rethinking_2025, tang_understanding_2025}, and decrease idea diversity in collaboration~\cite{meincke_chatgpt_2025}. These limitations raise widespread concerns regarding authenticity, ownership,  craftsmanship, and creative agency~\cite{bomba2024choreographer, ippolito2022creative, guo_pen_2024, zhao_making_2025, wan2024felt, kim2024authors, holopainen2025infinity}.
This tension suggests a complementary path. Instead of remaining desk-bound LLM assistants dependent on limited text entry, one possible solution is to enhance the human-in-the-loop approach \cite{wu2022survey, chung2021human} with richer user and real-world context \cite{jones2023embodying, kortenbach2025relation} fed to LLMs by engaging with writers’ everyday encounters, emotions, and observations. By grounding support in diverse lived experiences, such systems may help preserve authentic voices while increasing idea diversity.

\subsection{Mobile, Wearable, and Heads-up Computing for In-Situ Writing \& Inspiration}
\label{sec:rw_wearable}

To access richer user and environmental context, recent research has integrated AI, especially LLMs, into mobile \cite{kim2020livesnippets, dragonas2024forensic}, wearable \cite{jhajharia_wearable_2014, cliff_wearable_2005, meshi2024gpt, yuan2025alice}, and heads-up computing~\cite{cai_pandalens_2024,zhao2023headsup} systems. Among them, AI-enabled smart glasses show promise due to their ability for ubiquitous, real-time assistance in everyday activities while keeping user attention on the physical world~\cite{zhao2023headsup, lu2020glanceable}, opening new ways to reconnect embodied inspiration with creative work. Beyond ubiquitous, context-independent writing~\cite{nebeling2016wearwrite, zhou2024glassmail}, recent systems offer low-effort, context-aware support: PANDALens co-captures travel moments with users and yields higher-quality blogs than smartphone recordings~\cite{cai_pandalens_2024, kim2020livesnippets}; while GazePointAR \cite{lee2024gazepointar}, G-VOILA \cite{wang2024G-VOILA}, and AiGet \cite{cai_aiget_2025} sense gaze and environmental context to provide situated knowledge. Prior work also shows how mobile/smart glasses AR tools scaffold creativity across domains, such as storytelling and 3D authoring~\cite{kim_motif_2015, li_motivational_2023, tong2024vistellar, lee2025imaginateAR, yin2025traveltales}. For example, Motif is a mobile application that allows users to construct video stories via expert-derived patterns \cite{kim_motif_2015}, while classroom studies show that AR-based instruction improves descriptive and innovative writing, reducing off-task behavior~\cite{li_motivational_2023}. 

\added{More broadly, this line of work resonates with a long-standing HCI interest in treating everyday experience as material for open-ended interpretation and creative practice. Cultural Probes, for example, use evocative artifacts and tasks to elicit situated and subjective accounts of people’s daily lives as resources for interpretation and design inspiration~\cite{gaver1999probe}. Whereas probe responses are typically interpreted retrospectively by researchers, recent mobile, wearable, and AR technologies can support more immediate, in-situ engagement with lived experience. Viewed alongside this broader tradition, \CRAFT{} investigates how wearable AI can help writers notice contextual details, explore associations, and transform elements of everyday experience into fiction.}

Nevertheless, transforming lived experience into fiction introduces distinctive challenges~\cite{cohn2000distinction}. First, it involves \emph{dual-world bridging}: unlike lifelogging that directly records scenes and feelings~\cite{cai_pandalens_2024}, it requires high cognitive complexity for writers to simultaneously perceive physical environments and reinterpret elements into evolving fictional worlds with different internal logic and narrative rules. Second, as a \emph{long-term, large-scale creative effort}, it demands maintaining character, plot, and thematic consistency~\cite{cohn2000distinction} across sessions separated by hours, days, or months. For example, a couple's relationship inspired by a writer's restaurant encounter must remain coherent when depicting their behaviors during a later park visit, requiring cross-session support beyond the simpler support in existing systems (e.g., generating independent emails~\cite{zhou2024glassmail}). Together, these challenges establish creative transformation for fiction writing as a distinct domain requiring exploration of its desirability, feasibility, and potential viability.

\section{STUDY OVERVIEW}
\begin{figure*}[h]
    \centering
    \includegraphics[width=1\textwidth]{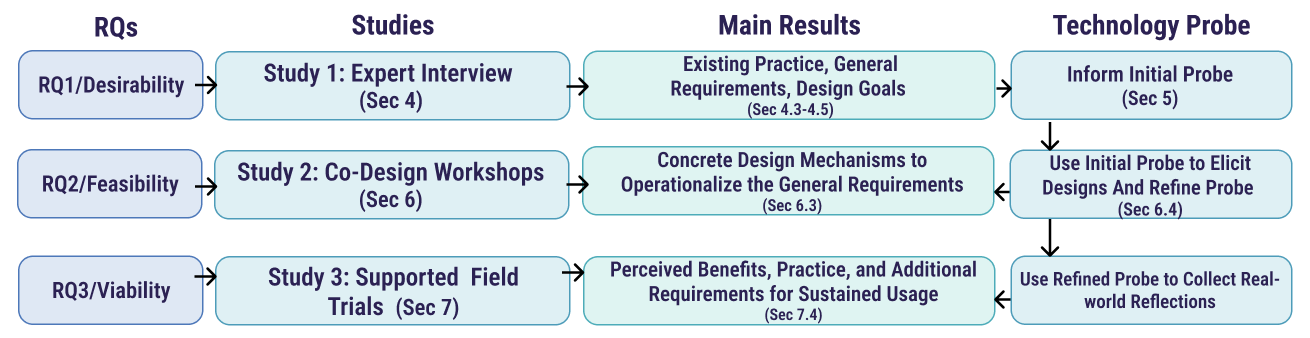}
    \caption{Relationship between RQs, studies, main results, and usage of technology probe.}
    \label{fig:study_overview}
\end{figure*}

As mentioned in the Introduction and illustrated in Figure~\ref{fig:study_overview}, our research comprises three complementary studies that progressively explore the \CRAFT{} approach's desirability, feasibility, and potential viability. All studies received approval from our university's Institutional Review Board (IRB), and participants were compensated at approximately USD 10 per hour.

\section{STUDY 1: EXPERT INTERVIEWS ON CREATIVE PRACTICE IN DAILY LIFE}
\label{sec:study1}

To understand the desirability of \wearableAI{}, we conducted interviews with creative and fiction writers. Our goals were to (G1) understand writers' current practices and challenges in transforming real-life experiences into fiction, and (G2) explore their attitudes and expectations toward AI and smart glasses for creativity support.

\subsection{Participants}

We interviewed 9 writers with published works or formal creative writing training in English/Chinese (5 female, 4 male, see Table~\ref{tab:study1_participants}). To capture diverse perspectives on traditional and AI-assisted writing, we recruited participants across two key dimensions: experience level and AI familiarity. Four senior writers (independent authors and literature professors with 30-46 years of experience) provided insights into established traditional practices, while younger participants offered perspectives on technology integration, including one with smart glasses experience. Participants' AI experience ranged from none to frequent daily use, and they represented diverse creative writing genres, including contemporary fiction, historical novels, poetry, prose, sci-fi, and fan fiction.

\begin{table}[h]
\centering
\small
\caption{Interview study participants: 9 writers with diverse experience levels and AI familiarity.}
\begin{tabularx}{\textwidth}{lllXlXX}
\hline
\textit{PID} & \textit{Gender} & \textit{Age} & \textit{Profession} & \textit{Experience (yrs)} & \textit{Main Genre} & \textit{AI Experience} \\
\hline
P1 & M & 65 & Independent Writer & 40+ & Contemporary \& historical novels & Occasional (information search) \\
\hline
P2 & M & 63 & Independent Writer \& Entrepreneur & 40+ & Contemporary fiction & Regular user but not for fiction writing \\
\hline
P3 & F & 23 & Amateur Fiction Writer \& Commissioned Writer & 10+ & Modern urban fiction, fan-fiction & Frequent for fiction content generation \\
\hline
P4 & F & 51 & Literature Professor & 30+ & Novel about historical writers, poetry analysis & Seldom \\
\hline
P5 & M & 68 & Literature Professor \& Professional Writer & 46 & Contemporary fiction & None \\
\hline
P6 & M & 31 & Book Writer \& PhD Student in Literature & 16 & Prose \& Creative Writing & Frequent for academic proofreading \\
\hline
P7 & F & 19 & Undergraduate in Literature & 2 & Campus fiction, fanfiction & Frequent for fiction details generation \\
\hline
P8 & F & 24 & Master's degree in Creative Writing with various Literary Awards & 6 & Short Fiction, Poetry, Prose & Frequent for getting feedback on writing\\
\hline
P9 & F & 37 & Independent Writer & 15 & Sci-fi and fantasy & Frequent for information search, but not for writing \\
\hline
\end{tabularx}

\label{tab:study1_participants}
\end{table}

\subsection{Methods}
\label{sec:study1_method}

We conducted 60-minute semi-structured interviews exploring current practices for capturing and transforming real-life experiences into fiction, challenges in this transformation process, and attitudes toward AI assistance and smart glasses for in-situ creativity support. To ensure informed opinions, we showed demonstration videos of existing AI smart glasses (e.g., Orion AI Glasses) before discussions. We performed inductive thematic analysis adapted from Braun \& Clarke's approach~\cite{braun2006thematic, braun2021thematic} to analyze interview transcripts (see Appendix~\ref{appendix:study_analysis} for details).

\subsection{Findings: Writers' Practices \& Challenges of Reality-Fiction Transformation in Daily Life (G1)}
\label{sec:study1_findings_principles}

All participants confirmed that real-world experiences and fiction creation are tightly coupled. While drafting primarily occurs in desktop environments, ideation, planning, and research often happen during daily activities. Table~\ref{tab:reality_fiction_practices} summarizes their practices. Writers engage in transformation both intentionally~\cite{biggs_fabulating_2025} (deliberately seeking environments) and unintentionally~\cite{williams_writer_2013} (spontaneous ideation during daily activities). They use environments, observed people, and personal experiences \cite{truelife_fassin_2014, zhao_making_2025} reflecting personalization, then transform them through selection and recombination, aesthetic processing (e.g., ``beautification,'' defamiliarization~\cite{shklovsky2015art}), and identity obfuscation for aesthetics and privacy.

\begin{table*}[htbp]
\caption{Writers' practices of reality-fiction transformation in daily life.}
\label{tab:reality_fiction_practices}
\small
\renewcommand{\arraystretch}{1.1}
\begin{tabular}{p{1.5cm}p{3.5cm}p{10cm}}
\toprule
\textbf{Aspect} & \textbf{Category \& Description} & \textbf{Examples} \\
\midrule
\multirow{9}{*}{\parbox{1.5cm}{\textbf{When:} \\ \scriptsize Creation opportunities in daily life}} 
& \textit{Intentional}~\cite{biggs_fabulating_2025} \par 
  \scriptsize Writers deliberately seek rich environments for conscious ideation. 
& \begin{itemize}[nosep, leftmargin=1em, label=\textbullet]
    \item Following ancient poet paths to imagine himself as the main character to provide immersive feelings for writing (P1)
    \item Taking \quote{1000-2000 notes by walking around and observing daily} before writing (P5)
    \item Intentionally collecting materials from life using notebooks and phone cameras, maintaining a digital repository for scattered raw materials (P6)
  \end{itemize} \\
\cmidrule{2-3}
& \textit{Unintentional}~\cite{williams_writer_2013} \par 
  \scriptsize Spontaneous ideation/inspiration during daily activities.
& \begin{itemize}[nosep, leftmargin=1em, label=\textbullet]
    \item Walking dogs, casual walking (P1, P8)
    \item Traveling (P3, P4, P6, P9)
  \end{itemize} \\
\midrule
\multirow{12}{*}{\parbox{1.5cm}{\textbf{What:} \\ \scriptsize Utilized elements from life as narrative components \cite{truelife_fassin_2014, zhao_making_2025}}} 
& \textit{Environment, Atmosphere, and Associated Emotion} \par 
  \scriptsize Draw on realistic environments and overall \quote{vibe} of places for fictional scenes.
& \begin{itemize}[nosep, leftmargin=1em, label=\textbullet]
    \item Google Maps Street Exploration or Real Seed Vault visit for sci-fi settings (P9)
    \item Leverage \quote{smoky, life-filled atmosphere} from travels (P3)
    \item Capturing \quote{objects, imaginations, emotions—three things} during storm (P6)
  \end{itemize} \\
\cmidrule{2-3}
& \textit{Observed People} \par 
  \scriptsize Encountered people's appearance, behaviors and dialogues inspire \quote{believable characters} (P1, P2).
& \begin{itemize}[nosep, leftmargin=1em, label=\textbullet]
    \item \quote{A one-armed grandmother skillfully weaving flower lanterns} (P3)
    \item Conversations between a young man and the woman he was pursuing that \quote{a writer can't even imagine, as dramatic plots must be drawn from life} (P5)
  \end{itemize} \\
\cmidrule{2-3}
& \textit{Personal Experience} \par 
  \scriptsize Fuse personal memories with narrative to make it reflect personalization.
& \begin{itemize}[nosep, leftmargin=1em, label=\textbullet]
    \item Use author's childhood cattle-raising experience for protagonist, creating \quote{a layer of truth that pure research could not provide} for historical fiction (P1)
    \item Job hunting experiences and feelings for contemporary novel (P8)
  \end{itemize} \\
\midrule
\multirow{14}{*}{\parbox{1.5cm}{\textbf{How:} \\ \scriptsize Techniques to transform raw materials into fiction ~\cite{boden2004creative} and make fictionalization~\cite{iser1990fictionalizing}}} 
& \textit{Selection \& Recombination} \par 
  \scriptsize Ensuring new elements align with narrative consistency.
& \begin{itemize}[nosep, leftmargin=1em, label=\textbullet]
    \item Keyword-based documentation systems to catalog and connect fragments (P5, P6)
    \item \quote{Cut and paste} traits from multiple people into composites with \quote{similar underlying behavioral logic} (P2)
  \end{itemize} \\
\cmidrule{2-3}
& \textit{Aesthetic \& Narrative Processing} \par 
  \scriptsize Adapting raw reality for artistic expression.
& \begin{itemize}[nosep, leftmargin=1em, label=\textbullet]
    \item Reality is \quote{beautified} to remove \quote{awkwardness} or \quote{slow tempo} (P3, P7): P7 \quote{beautified character personalities} in campus novel, viewing writing as a process to \quote{heal oneself.} 
    \item \quote{Defamiliarization}~\cite{shklovsky2015art} for \quote{uniqueness and uncommon expression} (P6, P7)
    \item Adapting narrative style to modern readers' taste (P4, P6)
  \end{itemize} \\
\cmidrule{2-3}
& \textit{Obfuscation for Privacy} \par 
  \scriptsize Changing names, jobs, cities.
& \quote{Good, real-life details are too precious to discard,} so making fictionalization, but \quote{people familiar with the real person can still recognize} (P5)
\\
\bottomrule
\end{tabular}
\end{table*}

\subsubsection{The Authenticity Triad}

\label{sec:study1:results:practice:authenticity}

These practices reflect a shared goal of \quote{making fictional worlds believable.} To sustain this sense of authenticity, writers described three layers of \quote{truth}, echoing prior discussions of factual accuracy, verisimilitude, and authenticity in fiction~\cite{doleel_truth_1980, saxton_true_2020, bruner1986actual, hannula_fictionalising_2003}:

\textbf{(1) Factual Accuracy:} Verifiable facts and dates provide anchors for reality-grounded stories. P1/P4 maintain clear distinctions for \quote{Historical Truth,} while P3/P8 emphasize the correctness of industry-specific terms and descriptions.

\textbf{(2) Logical \& Behavioral Consistency:} P2 emphasized preserving \quote{behavioral logic} of characters, while P3 focused on \quote{reasonable human activity sequences} and \quote{internal logical coherence} to avoid \quoteby{P8}{narrative bugs.}

\textbf{(3) Emotional \& Psychological Authenticity:} This ensures stories resonate with readers through \quote{Human Nature Truth} (P1)—psychologically believable actions and motivations—and \quote{emotional/sensory resonance} of experiences (P3).

\subsubsection{Current Challenges} 
Despite in-situ creation's benefits, writers face significant time and cognitive barriers: (1) \textit{Time limitations} (P1, P2, P3, P5), (2) \textit{Lack of suitable ubiquitous tools} leading to memory fade (P1), (3) \textit{Gaps between thought speed and output} (P2, P6), and (4) \textit{Hard to connect fragmented notes} (P8). P2 finds \quote{fragmented time} disrupts his ability to focus on turning daily inspirations into writing, often using drawing instead, as \quote{it's faster for capturing ideas.} P6's primary frustration is that his \quote{thoughts jump too fast,} and typing cannot keep up, making fleeting ideas lost.

\vspace{2mm}
In summary, these findings on current writers' practices and challenges inform when, what, and how \wearableAI{} can assist in the reality-to-fiction transformation, and guide the design direction to overcome writers' current limitations, which is detailed in the next section.

\subsection{Findings: Attitudes and Expectations for \wearableAI{} (G2)}
\label{sec:study1_findings_attitudes}

\subsubsection{Current AI Limitations and Opportunities}
\label{sec:study1:results:attitude:limitation}

Participants exhibited \textit{pragmatic but critical adoption} of current AI tools, ranging from heavy reliance (P3) to restricting them to non-creative tasks (P2, P9). Nevertheless, there was strong consensus on a core limitation: desktop AI lacks authors' grounded perception of the world (P2) and the ability to create character ``thickness'' (P3). As P7 noted, AI outputs often feel \quote{formulaic} and \quote{AI-like}, failing to evoke genuine emotion or authenticity rooted in real life. These concerns align with recent studies on AI-assisted creative writing~\cite{guo_pen_2024, lee_empirical_2024, meincke_chatgpt_2025}.

These limitations indicate opportunities for \wearableAI{} to capture \textbf{``authenticity triad'' materials from the real world, rather than relying \quoteby{P9}{solely on LLM training data} with authors' limited textual input.}

\subsubsection{Desired Functions and Boundaries for In-Situ Creation}
\label{sec:study1:results:attitude:desires}

All participants desired wearable AI to reduce creation time and effort, and we identified desired functions and their necessary boundaries, as detailed in Table~\ref{tab:desired_functions}.

\begin{table*}[htbp]
\caption{Desired functions and boundaries for wearable creative AI in fiction writing.}
\label{tab:desired_functions}
\small
\renewcommand{\arraystretch}{1.1}
\begin{tabular}{p{3.3cm}p{3.2cm}p{8.5cm}}
\toprule
\textbf{Category} & \textbf{Subcategory} & \textbf{Description \& Evidence} \\
\midrule

\multirow{10}{3cm}{\textbf{Augment Perceptions} \newline \scriptsize Capture fleeting moments and augment reality with creative intent} 
& \multirow{4}{3.5cm}{\textit{Augmenting Reality with ``Reality''} \newline \scriptsize Bridge knowledge \& memory gaps} 
& \textbf{In-Situ Research:} Real-time factual and historical research about locations or objects to \quote{spark associations and add realism} (P1, P3, P6). \\
& & \textbf{Past Memory Retrieval:} Proactive reminders by retrieving relevant past memories to \quote{bridge the memory gap} (P2, P7). \\
\cmidrule{2-3}

& \multirow{5}{3.5cm}{\textit{Augmenting Reality with ``Fiction''} \newline \scriptsize Overcome imagination gaps} 
& \textbf{Visualizing Fiction:} Show transformed fictional scenes on glasses, particularly helpful for sci-fi to visualize speculative elements (P3, P9). \\
& & \textbf{Associative Sparking:} Spark new connections between captured scenes and fictional content through AI-generated associations and ask \quote{what if} questions (P3, P8). \\
\midrule

\textbf{Direct Reality-Fiction Transformation}\newline\scriptsize  Enable faster iteration across possibilities 
& \textit{Drafting Assistance}\newline\scriptsize Address time barrier
& \quote{Imagine and fill in the blanks} for fictional scenes by transforming observations into fiction snippets. P7 noted: \quote{It can help you present various possibilities and perform these assemblies quickly and clearly.} \\
\midrule

\multirow{9}{3.6cm}{\textbf{Interaction Boundaries} \newline \scriptsize Ensuring enjoyment \& well-being~\cite{hwang_it_2025, gero_social_2023}} 
& \textit{Preserving Creative Agency \& Enjoyment} 
& \textbf{Authorial Authority:} Writers must remain the \quote{main entity} with the power to engage with, ignore, or edit suggestions (P1, P3). \newline\textbf{Enjoyment from Process:} Besides serendipity and reduced efforts from AI, enjoyment can stem from the \textit{process} of \quote{transforming ideas in mind into vivid content} (P3), not just the result (P8, P9)~\cite{paton_dynamic_2016, kim2024authors}. This writing process also provides emotional healing value (P6, P7). \\
\cmidrule{2-3}

& \textit{Life-Art Boundaries} 
& \textbf{Avoid Exhaustion:} Deliberate control (e.g., only for \quote{inspiration trips}, P3) is needed to prevent exhaustion from constant creative focus. \newline
\textbf{Trauma Avoidance:} Prevent \quote{trauma-informed} triggers~\cite{chen2022trauma} where AI might surface painful memories in daily life (P3). \\

\bottomrule
\end{tabular}
\end{table*}

(1) Regarding \textit{Functions}, beyond easy scene and feeling recording~\cite{cai_pandalens_2024}, participants desired two main functions: 1) \textit{Augmenting Perceptions} to bridge cognitive gaps in knowledge, memories, and imagination, and 2) \textit{Direct Reality-Fiction Transformation} to address time barriers (Table~\ref{tab:desired_functions}).
Crucially, they strongly preferred \textbf{in-situ support} for these functions: \quoteby{P7}{I think giving it right in the moment is best because my curiosity and desire to explore are at peak... If I wait... the emotions, the atmosphere, the connection to the past will all fade.}

(2) Regarding \textit{Boundaries}, writers insisted on remaining  the \quote{main entity} for creation control (P1, P3). This aligns with human-centered creative AI principles~\cite{hwang_it_2025,mccormack2020design}, driven by two motivations:
1) \textit{Preserving Creative Agency and Enjoyment}: writers valued the creative process itself for emotional fulfillment and rejected full automation~\cite{paton_dynamic_2016, kim2024authors}; 2) \textit{Maintaining Life-Art Boundaries}: they required system control to avoid constant creative pressure and trauma triggers~\cite{chen2022trauma}. These highlight the needs of context-aware, non-intrusive interactions that respect user agency and real-time intentions~\cite{zhao2023headsup, cai_pandalens_2024, cai_aiget_2025, lu2020glanceable}.

\subsection{Design Goals}
\label{sec:study1:design_principle}

From these findings, we derive three core design goals for \wearableAI{}:

\textit{\DPone{}: Augment in-situ perception to bridge reality and fiction}: \wearableAI{} should analyze multimodal context~\cite{cai_aiget_2025}, including real-time and historical user, environment, and fictional contexts, to provide contextual suggestions and ideation support, bridging the identified cognitive gaps.

\textit{\DPtwo{}: Promote authenticity triad while maintaining a degree of fictionalization}: When transforming content, it should sense the real world and author's emotional expressions while preserving a degree of fictionalization for aesthetics and privacy considerations through author confirmation or logical inference, addressing the authenticity triad we identified.

\textit{\DPthree{}: Preserve creative agency, enjoyment, and boundaries in daily life}: It should enhance enjoyment while balancing minimal explicit input with writers’ desire for expression, and reduce distraction and cognitive load~\cite{cherry_csi_2014}.

\section{TECHNOLOGY PROBE}
\label{sec:tech_prob}

Study 1 confirmed writers' desirability of the \CRAFT{} approach for fiction creation. To explore its concrete design and emergent practices (RQ2 \& RQ3), we adopted a technology probe methodology~\cite{hutchinson2003technology, huang2014technology}. Serving to elicit user needs and inspire reflection rather than usability testing~\cite{huang2014technology, yin2025traveltales, barbosa2021ads, gaver1999probe}, the probe offered open-ended, minimal core functions for in-situ fiction writing. We implemented an \textit{Initial Probe}—reflecting the design goals and requirements from Study 1—to provoke concrete design feedback in Study 2's co-design workshops. Incorporating these insights, we evolved it into a \textit{Refined Probe} for the multi-session field trials in Study 3. The following section details the design of the \textit{Initial Probe}.

\begin{figure}[ht]
    \centering
    \includegraphics[width=1\linewidth]{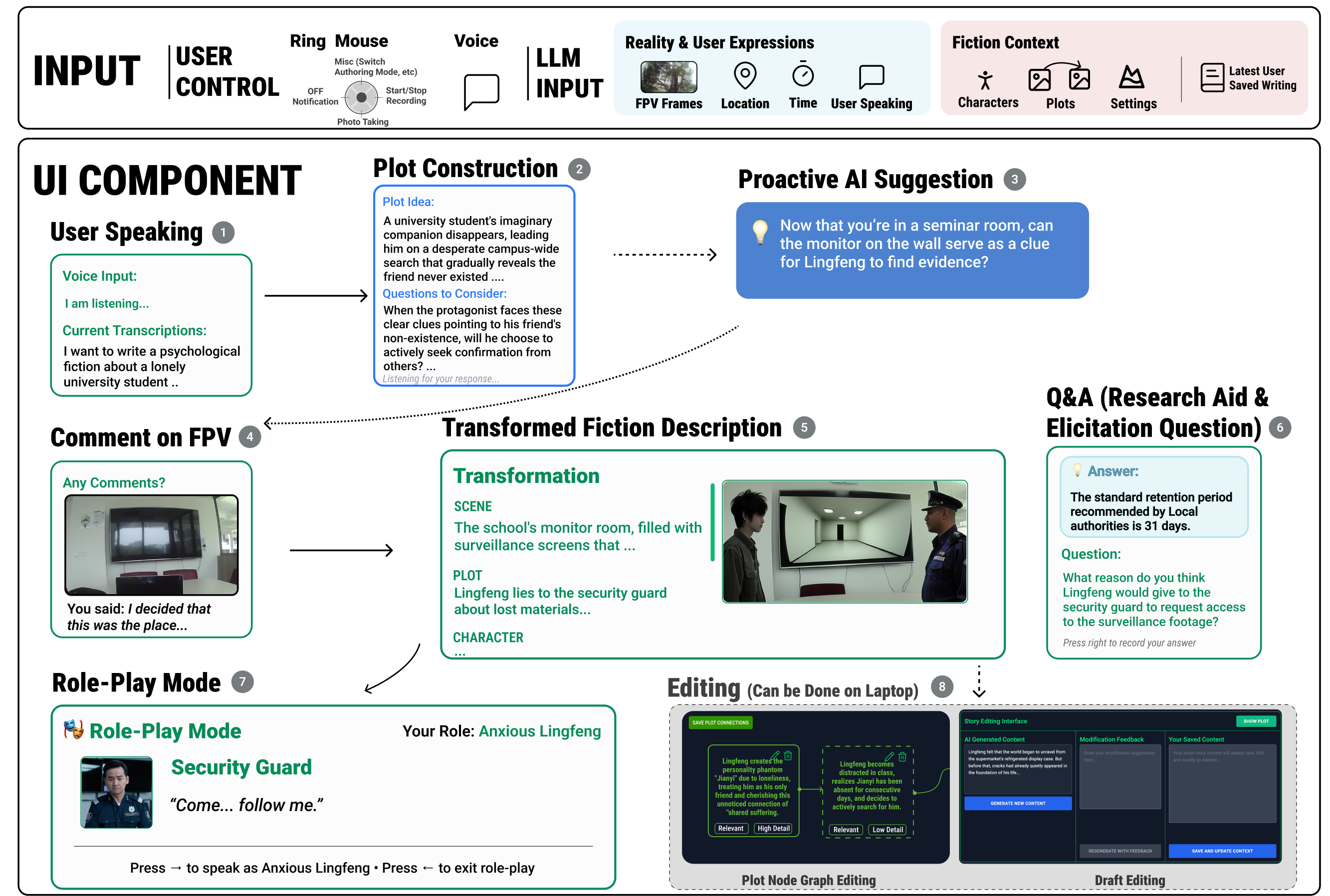}
    \caption{\CRAFT{} probe overview. Smart glasses capture first-person video, speech, time, and location. Using mixed-initiative interaction—user-initiative ((1) comments or (4) photos) and AI-initiative (continuous sensing)—the probe fuses these data with story context to deliver on-glasses (2) plot construction, (3) proactive suggestion, (5) transformed fiction description, (6) Q\&A, and (7) role-play. It also supports (8) laptop editing after in-situ creation. Note: Role-Play mode was added after co-design workshops and used in supported field trials. The color in Smart Glasses UI is slightly modified to ensure visibility for printing.}
    \label{fig:design_probe}
\end{figure}
\subsection{Mixed-Initiative Interaction Workflow}
The probe adopts a mixed-initiative approach balancing AI proactivity with writer control \cite{allen1999mixed-initiative_interaction, horvitz1999principles}, aligned with our design goals.

\subsubsection{Plot Construction: Setting Initial Context} Writers begin by establishing fiction's theme, topics, and style through voice (Figure~\ref{fig:design_probe}-1,2), with the probe asking elicitation questions to guide ideation. This creates an initial global fictional context that shapes subsequent transformations. Writers can review and revise these global elements at any later stage.

\subsubsection{AI-Initiative: Proactive Suggestions (\DPone{})} 
To augment in-situ perception, the initial probe proactively suggests creation opportunities within the immediate environment  (Figure~\ref{fig:design_probe}-3). By continuously analyzing real-world context (e.g., objects, atmosphere, and people, following sec~\ref{sec:study1_findings_principles} and \ref{sec:study1:results:attitude:desires}), via user location and First-Person-View (FPV) camera streams, the probe utilizes a Multimodal Large Language Model (MLLM) to detect semantic associations between physical surroundings and the established fictional context. This mechanism bridges reality and fiction; for instance, as a writer working on a café scene passes a physical coffee shop, the probe proactively suggests relevant atmospheric details or potential plot connections.

\subsubsection{User-Initiative: Photo \& Voice-Based Reality Transformation (\DPone{}, \DPtwo{})} 
When inspired—either by AI suggestions or independently—writers can capture observations via the ring mouse accompanied by voice commands (Figure~\ref{fig:design_probe}-4). An MLLM then synthesizes the multimodal context (FPV frames, time, location, user comments, and prior fiction context) into authentic fictional scenes, plots, or characters. The output is presented via AR text overlays and audio feedback (Figure~\ref{fig:design_probe}-5). 
To support dual-world awareness, the probe system generates \emph{half-real/half-fictional} visualizations: hybrid images where only specific fictional elements are superimposed onto the captured real-world photo. To support the authenticity triad, the probe identifies knowledge queries within user comments and presents factual answers alongside the creative transformation (Figure~\ref{fig:design_probe}-6). Finally, the system poses follow-up questions to facilitate deeper ideation and local refinement, continuously updating the global fiction context.

\subsubsection{Full Draft Generation \& Editing (\DPtwo{})} 
After in-situ sessions, writers can transition to a desktop interface for full draft generation (Figure~\ref{fig:design_probe}-8). The system aggregates the accumulated fiction context—comprising characters, plots, settings, and interaction history—into a \textit{plot node graph}~\cite{qin2025node} to facilitate global narrative management. An LLM then synthesizes these elements to produce or expand the draft, ensuring narrative consistency across distributed sessions.

\subsubsection{Interface Design for Reducing Daily Activity Interference (\DPthree{})}
To preserve creative boundaries, the UI utilizes \textit{peripheral positioning}~\cite{janaka2022paracentral, cai_pandalens_2024, cai2023paraglassmenu} to minimize distraction from the primary task (e.g., placing proactive suggestions in the upper-right peripheral visual field to balance visibility with unobtrusiveness). To mitigate intrusiveness, these proactive suggestions appear visually for 30 seconds without audio cues. Writers retain agency via ring-based interaction~\cite{sapkota2021ubiquitous}, allowing them to dismiss outputs at will. 
To further reduce information overload, the probe also limits repetitive proactive suggestions through redundancy control (detailed in sec~\ref{sec:tech_prob:implementation} and Appendix~\ref{appendix:proactive_pipeline}).

\subsection{Apparatus and Implementation}
\label{sec:tech_prob:implementation}

The probe uses Xreal Air glasses connected to a MacBook Pro (M2 Pro) with a Sanwa Bluetooth ring mouse for ubiquitous and subtle input~\cite{sapkota2021ubiquitous}, and a Pupil Core camera capturing FPV frames. The software utilizes a React frontend and a Python backend, with Gemini models for multimodal analysis and generation, OpenAI TTS for audio synthesis, and few-shot prompting~\cite{brown2020language} to guide transformations. To balance responsiveness and cognitive load, the proactive pipeline runs on a 30-second interval loop based on pilot testing. To reduce repetition and distraction, beyond instructing the MLLMs to avoid generating suggestions similar to recent history, the \emph{Initial Probe} employs a two-stage filter. First, a MobileNet~\cite{howard2019searching}-based visual pre-filter skips visually static or repetitive FPV contexts using both frame-level and window-level similarity checks. Second, a textual semantic post-filter using all-MiniLM-L6-v2~\cite{wang2020minilm} suppresses generated suggestions that are similar to recent suggestion history. \added{The triggered MLLM pipeline latency averaged 8.14 seconds (SD=0.90) for proactive suggestions and 8.95 seconds (SD=1.01) for user-initiated queries during pilot testing}, close to the 9-second condition reported in prior work~\cite{tan2026impact} as a balanced delay between perceived AI thoughtfulness and frustration in creation-oriented human--LLM interaction. The probe maintains synchronized contexts (user preferences, environmental data, fictional elements) across sessions for narrative consistency. Full implementation details are provided in Appendix~\ref{appendix:design_probe}.

\section{STUDY 2: CO-DESIGN WORKSHOPS}
\label{sec:study2}
To operationalize the general requirements from Study 1 into concrete interaction mechanisms (RQ2-Feasibility), we conducted co-design workshops with writers and HCI/AI experts. Participants engaged with the \textit{Initial Probe} to ground their experience, utilizing the design goals as stimuli to concretize interaction mechanisms across everyday contexts. 
These insights also informed the probe refinement described at the end of this section.

\subsection{Participants}
\label{sec:study2_participants}

We conducted four workshops with 16 participants (8 female, 8 male, age range: 22--42). Each workshop included 2 creative writers and 2 researchers to foster a rich, interdisciplinary dialogue. 
Specifically, eight participants with a writing background hold literature-relevant degrees, pursue writing as a career, or have publications, with a mean of 7.0 years of active creative writing experience (SD=5.8 years). Seven had fiction writing experience, and one had screenwriting experience. Their familiarity with using AI in creative writing varied (range: 1-5, \meansd{3.1}{1.5} out of 5). Another eight participants who have a research background include six who specialize in wearable devices (smart glasses) and two in AI/Human-AI interaction. They served as novice writers who enjoyed reading fiction.

\subsection{Procedure}
\label{sec:study2_method}
\begin{figure}[h]
    \centering
    \includegraphics[width=1\linewidth]{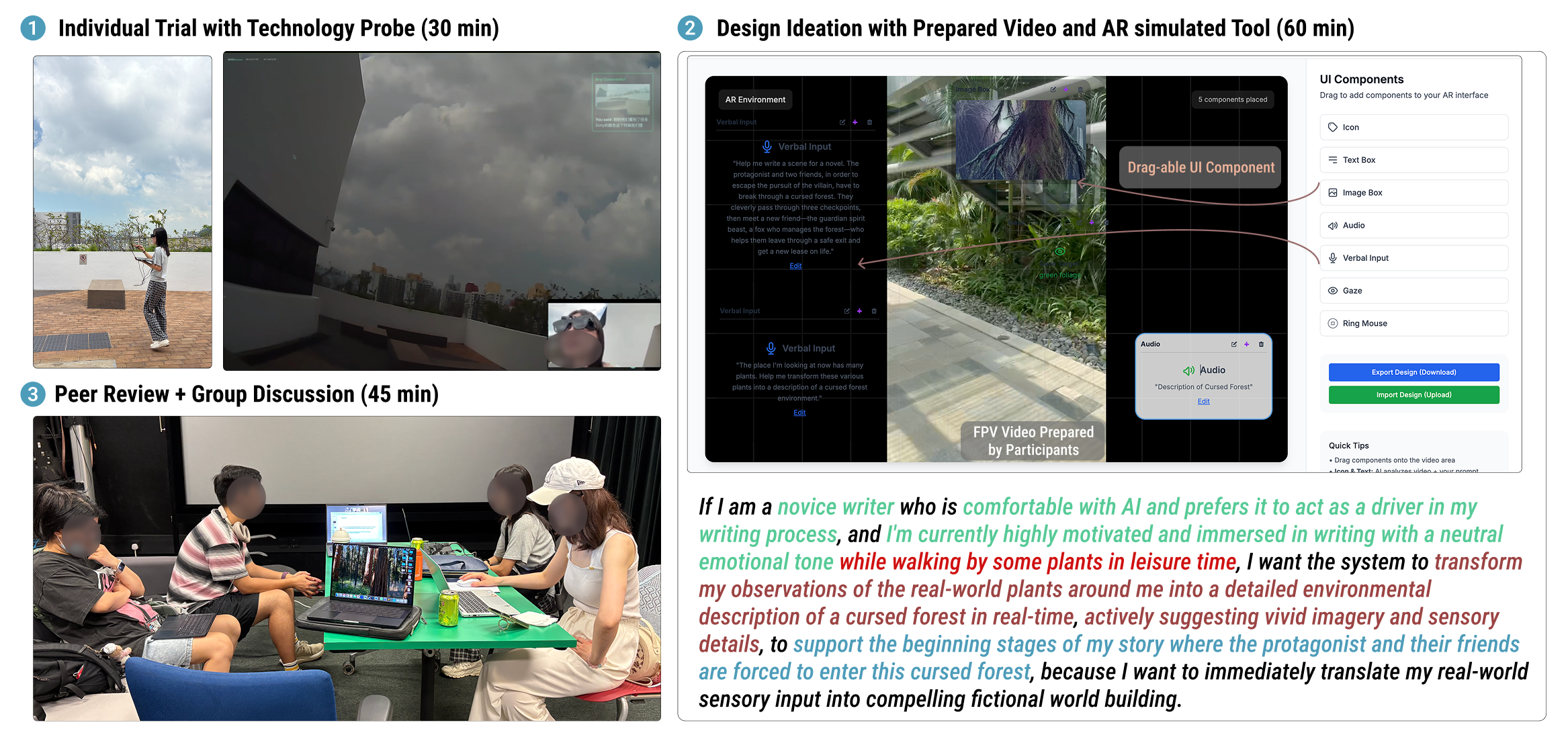}
    \caption{Design workshop procedure: (1) technology probe trial, (2) design ideation with a simulated AR tool where users can drag UI components (Input: verbal, ring mouse, gaze; and Output: text, image, audio, icon) as overlays onto their prepared FPV videos. The AR simulation tool enables manual editing of output design or AI-assisted content generation by analyzing the video frames and user-set context. Users also write design explanations or specify alternative designs not provided by the tool, and (3) group discussion.}
    \label{fig:study2_procedure}
\end{figure}

Our workshops progressed from hands-on experience to design ideation (Figure~\ref{fig:study2_procedure}). Before the session, each participant prepared two short daily-life videos: one from a routine setting and one from a less-frequented place~\cite{williams_writer_2013}. The workshop then consisted of three phases. First, in \textit{Introduction \& Probe Experience} (30 min), participants tried the initial probe (sec~\ref{sec:tech_prob}) on campus to create brief fiction snippets and gain a grounded understanding of its current capabilities~\cite{huang2014technology}. Second, in \textit{Design Ideation} (60 min), after reviewing the initial design goals, participants used a simulated AR tool to design 3--4 desired interaction mechanisms by placing input/output UI components on frames from their videos, and wrote brief rationales describing the context, writing goals, desired function, and use case. Third, in \textit{Peer Review \& Group Discussion} (45 min), participants reviewed one another's designs and consolidated insights. Workshops were recorded with audio and video in Zoom Cloud.

\subsection{Findings}
\label{sec:study2_findings}
By synthesizing probe interaction logs, proposed designs, and group discussions, we identified three key themes that concretize and extend the desired interactions and design goals from Study 1: (1) \textit{what} reality entities to prioritize for transformation; (2) \textit{what} forms to transform them into; and (3) \textit{how} to preserve creative boundaries and enjoyment.

\subsubsection{What to Transform: Reality Entity Prioritization (\DPone{}\&\DPtwo{})}
\label{sec:study2:findings:what_to_transform}
\begin{figure}
    \centering
    \includegraphics[width=1\linewidth]{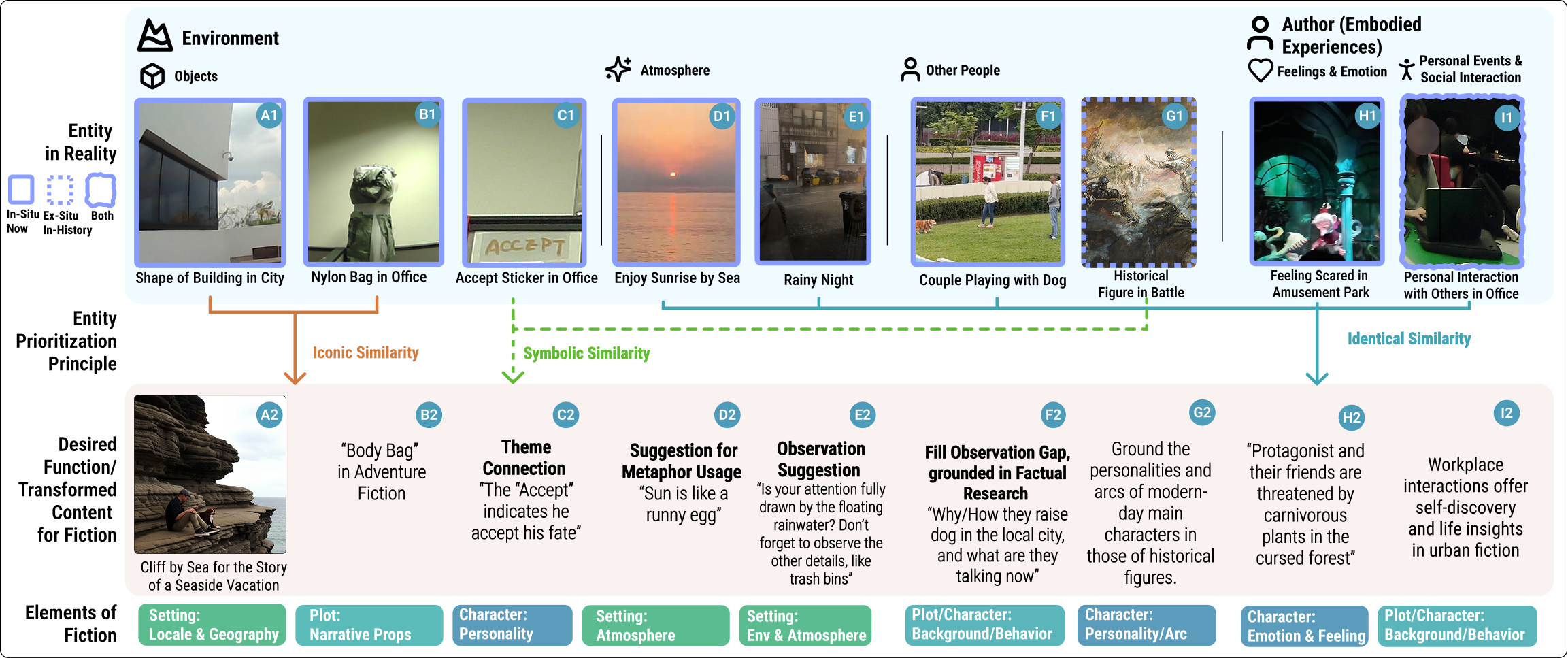}
    \caption{Workshop examples of reality-fiction transformation. Real-world entities (top) map to fictional elements—setting, plot, and character—via three similarity relations (\emph{identical}, \emph{iconic}, \emph{symbolic}).}
    \label{fig:workshop_examples}
\end{figure}

While Study 1 categorized the general landscape of utilizable resources (Table~\ref{tab:reality_fiction_practices}-\textit{What}), prior research on proactive AI emphasizes that identifying task-relevant entities is critical when operating in complex environments~\cite{cai_aiget_2025}. To address this challenge, we identify three \textit{similarity relations} that serve as detection criteria for \wearableAI{}. These relations are grounded in Peirce’s theory of signs~\cite{peirce1985logic}—adapted through our workshop cases (Figure~\ref{fig:workshop_examples})—and corroborate the writer practices observed in Study 1.

\begin{itemize}
    \item \textit{Identical similarity:} A real-world entity maps directly to its fictional counterpart, preserving authenticity through one-to-one mirroring. For example, P7 transformed an observed rainy night directly into a fictional rainy night. Notably, identical transformations can also carry \emph{indexical} (cause-and-effect) relationships to other fictional elements \cite{peirce1985logic}. P11, for instance, described a laptop in the lab as a laptop in the story, which then implied a plotline where a cat character attempts to use the computer to build and edit a genetic database of humans.
    \item \textit{Iconic similarity:} Abstraction of salient features (shape, texture, function) to fuel imagination. For example, P2 abstracted the facade of a modern building into the cliff of a seaside scene, while P4 transformed a \quote{nylon bag} into a \quote{body bag} in the fiction. This balances authenticity with creative abstraction.
    \item \textit{Symbolic similarity:} Symbolic/cultural meanings linked to broader themes. For example, P3 described how a wall sign reading “Accept” echoed a character’s fate in commercial conflicts. She also drew on personalities from ancient epics to model character arcs, using symbolism to preserve authentic human logic.
\end{itemize}

These relations operationalize entity prioritization, enabling future systems to: (1) identify high-potential items for proactive suggestions (\DPone{}), and (2) facilitate direct transformation into fictional elements (\DPtwo{}).

\subsubsection{Transformed into What: Multi-Layered Perceptual Augmentation (\DPone{}\&\DPtwo{})}
\label{sec:study2:findings:tranformed_into_what}
The participants also specified how real-world entities can be better utilized. From their proposed design, we synthesized a multi-layered perceptual augmentation that processes reality in three ways: it can broaden the writer's observational scope, deepen their cognitive understanding through authenticity grounding, and translate raw sensations into unique literary expressions.
\begin{itemize}
\item \textit{Augmenting Perceptual Breadth: Attention Steering \& Bias Mitigation.}
\label{study2:results:what:observation_guide}
Participants wanted \wearableAI{} to act as an observation coach, prompting them to notice \quoteby{P14}{overlooked elements in the environment} and \quoteby{P7}{countering observational bias}~\cite{mack2003inattentional}. For instance, while writing about a rainy night, P7 wanted the AI to remind her to observe the \quote{original setup of this environment}—such as trash bins or building arrangements—rather than focusing solely on the rain.

\item \textit{Augmenting Perceptual Depth: Contextual Anchoring via Factual Grounding.}
\label{study2:results:what:factual}
Participants wanted AI to fill observation gaps by integrating factual research. For example, P2, after observing a couple with a dog in the park, hoped AI could generate plausible backstories or dialogue grounded in local policies (e.g., dog ownership rules). Similarly, P6 expressed a desire for AI to provide concrete details (species, appearance) when fleeting observations of a pigeon were incomplete.

\item  \textit{Augmenting Perceptual Translation: Transforming Sensation into Metaphor.} P2 sought assistance in transforming direct observations into richer fictional descriptions by generating metaphors (e.g., a \quote{sunrise like a soft-boiled egg}), thereby linking sensory impressions (e.g., \quote{hungry feeling}) to emotionally resonant narrative elements.
\end{itemize}

This multi-layered design informs the future system development: (1) suggesting overlooked details in \textit{proactive suggestions} (\DPone{}); and (2) integrating external knowledge and translating sensory data into proactive suggestions (\DPone{}) or directly transforming them into literary expressions (\DPtwo{}).

\subsubsection{How to Transform: Considerations for Boundaries \& Enjoyment (\DPthree{})}
\label{sec:study2_findings_how}

Participants also emphasized the importance of protecting creative flow and life-art boundaries while enhancing enjoyment when creating in situ, and provided specific designs, including adaptive interactions and immersive enhancements.

(1) \textit{Preserving Boundaries via Adapting Interaction.}
Participants emphasized adapting AI interaction \emph{timing} and \emph{intensity} to their situational context. In dynamic contexts (e.g., amusement parks), they prioritized situational awareness to \quote{focus on enjoying the moment} (P4), favoring lightweight or delayed outputs~\cite{cai_pandalens_2024}—such as postponed suggestions (P2), only presenting key suggestions (P3), or salient notifications for later review (P7). Conversely, relaxed settings invited prolonged, detailed interaction (e.g., brainstorming about plants). To manage this, participants proposed \textit{Configurable Proactivity Modes} (e.g., \quote{field research}, \quote{light/detailed suggestion}, or \quote{AI disabled} modes [P2, P3]) to tailor AI proactivity and output length/speed to their context and intentions.

(2) \textit{Enhancing Enjoyment via Immersive Authoring.} To enable convenient creation (e.g., for dialogue) and enhance enjoyment during authoring, participants also designed playful and immersive extensions beyond initial AR visualization of the scene, such as role-playing conversations~\cite{tang_understanding_2025} with fictional characters (P3). Similarly, P14 mentioned that the audio feedback could include not only scene descriptions but also engaging elements~\cite{de2025sonora} such as \quote{environmental sounds or reminders from characters.}

\vspace{2mm}

Together, these findings addressed \emph{RQ2-Feasibility} by providing concrete designs for \wearableAI{}.

\subsection{Probe Refinement}
\label{sec:probe_refinement}
Building on the workshops' insights, we refined the \textit{Initial Probe} for the following supported field trials. As a technology probe, rather than implementing all the new features suggested~\cite{li2023ar_story}, we selected the core interaction design that represents three identified themes from the workshops. Key updates included:
1) \textit{Context-Aware Proactive Suggestions:} incorporating similarity detection  (sec~\ref{sec:study2:findings:what_to_transform}) via few-shot MLLM prompts using workshop examples to determine identical, iconic, or symbolic opportunities, and supporting new suggestion types like observation reminders, gap filling, and metaphorical enhancement (sec~\ref{sec:study2:findings:tranformed_into_what}) via few-shot LLM prompts. Suggestion timing was also regulated by MLLM analysis of environmental context and user interruptibility (sec~\ref{sec:study2_findings_how}).
2) \textit{Role-Play Mode:} A new immersive feature (Figure~\ref{fig:design_probe}-7) allowing users to play one character while the AI plays another by generating the partner's image and voice. Users can click the ring mouse to enter this mode and use voice to reply, switch to a different character, or ask the AI to adjust the tone or expression if not satisfied. 
Implementation details of these updates are in Appendix~\ref{appendix:design_probe}.

\section{STUDY 3: SUPPORTED FIELD TRIALS OF THE REFINED PROBE}
\label{sec:study3}

To explore the \emph{potential} viability of the \CRAFT{} approach (RQ3), we conducted supported field trials with the refined probe (8 writers, 24 sessions). Rather than comparing against traditional writing baselines, we focused on understanding in-situ usage patterns, emergent creative practices, and long-term design requirements through participants' free-form interactions and post-session reflections. While a larger user base and an extended deployment would provide more comprehensive evidence, this exploratory study provides initial insights into how wearable creative AI might integrate into writers' everyday creative practice.

\subsection{Participants}
\label{sec:study3_participants}

Following the recruitment and stopping rules in Appendix~\ref{appendix:recruitment}, we recruited 8 participants (5 female, 3 male; aged 22-33; 2 English and 6 Chinese writers) from diverse backgrounds. Six had substantial creative writing experience (5 in fiction, 1 in film scripts; 4--22 years of active writing): three had book, magazine, and online publications, while two had only online fiction publications. Two hosted fiction writing communities, and two won fiction awards. To assess if \CRAFT{} reduces barriers for beginners, we also included two enthusiastic fiction readers with writing ideas but no prior writing experience. This allowed us to observe how a \wearableAI{} could support writers at different levels.

\subsection{Study Design and Procedure}
\label{sec:study3_design_procedure}
\begin{figure}[h]
\centering
\includegraphics[width=0.76\textwidth]{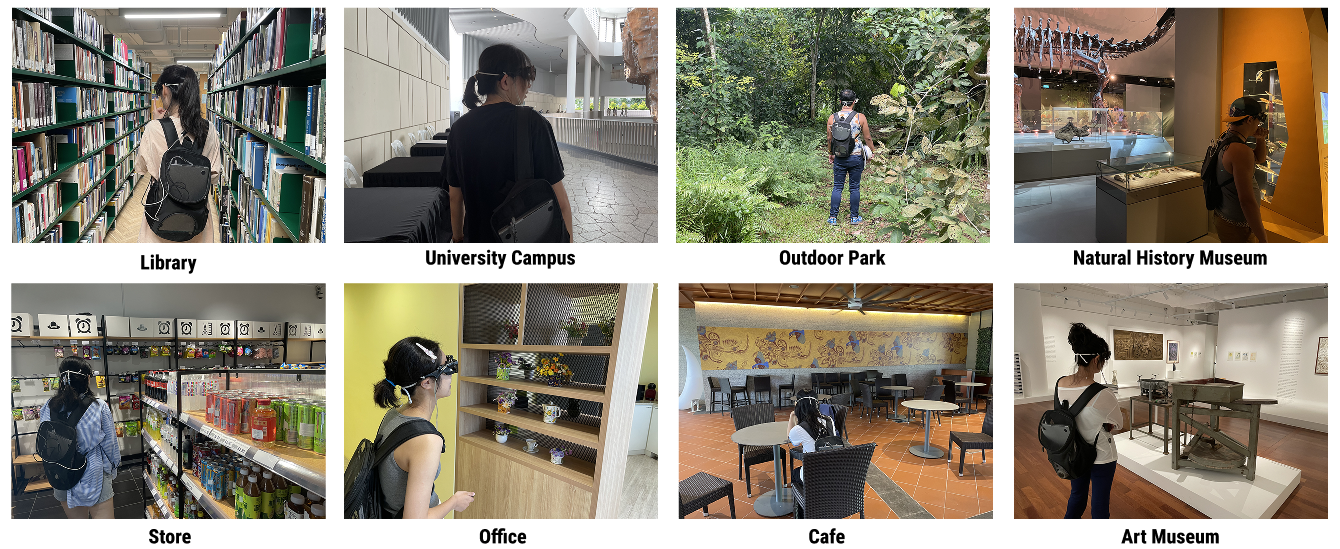}
\caption{Some locations for Study 3, selected based on participants' preferences and location availability.}
\label{fig:study3_locations}
\end{figure}

Following prior smart-glasses field study designs~\cite{tan_audioxtend_2024, cai_aiget_2025}, each participant completed three self-directed sessions in different locations over one week, totaling 24 sessions (Figure~\ref{fig:study3_locations}). This multi-session design allowed us to observe how creative practices evolved across contexts.

Participants first completed a training session. Each subsequent session followed three stages: (1) \textit{In-Situ Creation} (60 min), in which participants used the probe in a chosen environment while the experimenter observed from a distance to minimize interference; (2) \textit{Post-Session Revision and Interview} (30 min), in which participants reviewed recorded ideas and AI-generated drafts, optionally revised them, and then completed questionnaires and semi-structured interviews; and (3) \textit{Take-Home Revision} (optional), in which participants could further revise their work using the recorded ideas and AI suggestion history. A longer final interview (40 min) was conducted after the last session.

\subsection{Measures}
\label{sec:study3_measures}

To characterize participants’ experiences with the \CRAFT{} approach and corroborate interview feedback, we collected both the quality of the creation support and interaction from participants (detailed definitions are in Appendix Table~\ref{tab:study3_measures_ops}).

\subsubsection{Quality of Creation Support}

We adopted relevant measures from the Creative Support Index \cite{carroll_csi_2009, cherry_csi_2014} using 7-point Likert scales, including \exploration{}, \enjoyment{}, \expressiveness{}, \outputQuality{}, and \worthEffort{}, as descriptive indicators of author-perceived creative support following its usage scenarios~\cite{cherry_csi_2014}. We also collected the count of characters, scenes, plots, and words as descriptive statistics of the final narratives.

\subsubsection{Quality of Interaction} Drawing on prior work on wearable assistants for smart glasses~\cite{cai_pandalens_2024, cai_aiget_2025}, we measured \lowDistraction{}, \control{}, \unexpectedness{}, \unnoticedDetail{}, \useOfDailyObservation{}, and \deepenEnvironment{} using a 7-point Likert scale. We also recorded \taskDuration{} and interaction logs between users and AI.

\subsubsection{Data Analysis}
\label{sec:study3_analysis}

We conducted thematic analysis \cite{braun2006thematic, braun2021thematic} of interviews, observations, and artifacts to identify creative practices and requirements (see Appendix~\ref{appendix:study_analysis}) and quantitative analysis of ratings across sessions and participants.

\subsection{Results and Findings}

\subsubsection{Overall Experiences: Indications of Potential Viability}
\label{sec:study3:results:overall_exp}
We first summarize the study results before delving into detailed findings. 
The eight participants (P1--P8) each produced one short fiction spanning science fiction (3), detective/thriller (2), psychological suspense (1), speculative fiction (1), and historical fiction (1), with themes including implanted knowledge chips, alien dogs from university experiments, illusory friends, and revolutionary youth.

Each short fiction was developed through three one-hour in-situ writing sessions (\meansd{62.0}{12.5} minutes), totaling 24 sessions across 8 participants. The average length of each story is 4,076 words (SD=2105). Despite the time-constrained experience of crafting short stories, participants reported largely positive self-perceived outcomes, with \outputQuality{} at (5.9/7) and \worthEffort{} at (5.8/7) on average.

Beyond the author-perceived outcomes, the interaction logs provide further indicators of \CRAFT{}'s potential viability. As shown in Figure~\ref{fig:study3:usage_patterns}, across these sessions, 849 interactions were performed in total, through individual fictional element (53.4\%) and global plot (15.4\%) management, proactive suggestion (15.3\%), role-play dialogues (9.0\%), and Q\&A (6.9\%). Through these interactions, writers integrated 215 real-world elements into their fictional narratives: 143 objects (e.g., exhibits, furniture), 33 people (e.g., musicians, historical figures), and 39 settings (e.g., gardens, galleries). Individual sessions utilized between 3 and 39 elements (M=9.0), leading to 8.9 scenes (SD=5.1), 11.3 characters (SD=5.3), 11.1 distinct plot events (SD=3.3), and 14.4 character speeches or internal thoughts (SD=7.3) per final story. This suggests that participants used \wearableAI{} to develop contextually grounded story material from real-world observations.

We next examine this potential viability through: supporting factors from our design goals (DG1–3), usage patterns across diverse creative workflows, and requirements for sustained real-world usage.

\begin{table*}[htbp]
\caption{Factors informing future real-world viability: mechanisms and evidence.}
\label{tab:study3_unified_viability}
\footnotesize
\renewcommand{\arraystretch}{0.8} 
\begin{tabular}{p{2.7cm}p{4.2cm}p{8.9cm}}
\toprule
\textbf{Mechanism/Pattern} & \textbf{Description} & \textbf{Creation Evidence \& User Perspectives} \\
\midrule

\multicolumn{3}{l}{\textit{\textbf{1. In-situ Inspiration Support (\DPone):} Making the environment as active stimulus}} \\
\midrule
\textbf{Enhanced Idea Sources Noticing} 
& Filters noise and directs attention to environmental details matching narrative themes. 
& P2 (Dystopian Sci-Fi) received prompt linking street drummer's rhythm to \quote{signals of rebellion} $\to$ shifted passive observation into active idea of time signatures for coded messages. \\
\cmidrule{2-3}

\textbf{Real-time Capture of Fleeting Inspiration} 
& Preserves fleeting thoughts and details that fade before desktop documentation. 
& \quote{You walk somewhere, a thought hits—like, 'What if the kitchen hides a clue to the murder?'—and you tell the AI directly. If I had to wait to write it down, I'd forget the feeling of the moment, and the idea would die.} (P4). \\
\cmidrule{2-3}

\textbf{Serendipitous Adoption} 
& Reframes mundane, unplanned moments as narrative pivotal points. 
& P5 (Mystery): An accidental mosquito bite during testing $\to$ sparked plot of antagonist using controlled mosquitoes for blood collection. \\
\cmidrule{2-3}

\textbf{Contextual Knowledge} 
& Deepens surface observations with theoretical or historical context. 
& P5 (Mystery): Roadside flowers became inspiration when AI mapped them to the Chinese ``Five-Element'' theory $\to$ creating a color-coded conspiracy system in fiction. \\
\cmidrule{2-3}

\textbf{AR Visualization} 
& Bridges abstract imagination with tangible scale and physics. 
& P6 (Sci-Fi): Struggled to visualize \quote{pillar damaged in monster attack} until AR transformed campus pillar $\to$ specific damage pattern (Fig.~\ref{fig:img_examples}-1) led to imagining creature details. \\
\midrule

\multicolumn{3}{l}{\textit{\textbf{2. Reality Utilization in Creative Work (\DPtwo):} Transforming observations into authentic narrative elements}} \\
\midrule
\textbf{Identical Capture} 
& Integrating physical scenes and objects as concrete story elements. 
& P1 (Psychological Thriller): Used familiar campus locations to ground plot points, utilizing the real physical layout of a seminar room to structure the scene where the protagonist discovers imaginary friends: \textbf{Promotes Factual (campus layout) and Emotional (Personal familiarity) Authenticity} \\
\cmidrule{2-3}

\textbf{Iconic Transform} 
& Mapping visual form or structure from reality to fiction. 
& P4 (Murder): Transformed an art sketch of a man's back into a painting of a woman playing violin. The woman's leaning posture visually mirrored the murder victim's death pose: \textbf{Promotes Factual (art references) \& Emotional (Haunting irony) Authenticity} \\
\cmidrule{2-3}

\textbf{Symbolic Transform} 
& Mapping abstract meanings or cultural associations. 
& P3 (Speculative): Used local Kristang cultural symbols found in the environment to create metaphorical bridges connecting storylines across past, present, and future timeframes: \textbf{Promotes Factual (cultural references) \& Emotional (cultural identity resonance) Authenticity} \\
\cmidrule{2-3}

\textbf{Holistic Mapping} 
& Using multi-source real-world contexts under coherent knowledge/theory structures to produce consistent narratives. 
& P7 (Historical): Utilizing \textit{spatial logic} of a modern campus (dark rooms, wall etchings) with 1920s historical figures' true stories, ensuring character behaviors followed authentic constraints: \textbf{Promotes Factual (Historical accuracy) \& Logical (Character goals and relations match records) \& Emotional (Authentic period atmosphere)} \\
\midrule

\multicolumn{3}{l}{\textit{\textbf{3. Creative Enjoyment with Boundaries (\DPthree):} Balancing immersive embodiment and boundaries}} \\
\midrule
\multirow{4}{*}{\parbox{2.8cm}{\textbf{Immersive Authoring}}} 
& \textbf{1st-Person Embodiment:} Viewing the world through the protagonist's eyes to frame decisions \& emotions. 
& \quote{You're embodying that character... You walk around campus, thinking about what would happen to them} (P1). \\
\cmidrule{2-3}
& \textbf{3rd-Person Dialogue:} Role-play with AI characters; user acts as mediator or story inhabitant. 
& \quote{I felt like I was really talking to the historical figure by the lake... I stepped into the protagonist's shoes to converse with him} (P7). \\
\midrule

\multirow{5}{*}{\parbox{2.8cm}{\textbf{Trade-offs \& Boundaries}}} 
& \textbf{Psychological Impact:} Managing ``emotional bleed''~\cite{hugaas2024bleed} where intense genres affect real-world mood. 
& P1 noted dark thriller themes could affect their real-world mood. \quote{Like method acting, it will definitely affect your psychology.} \\
\cmidrule{2-3}
& \textbf{Distraction Handling:} AI interactions and the physical presence of the glasses could impose attentional demands, requiring design mitigations. 
& \quote{I can't ignore the tool's existence as it gives many suggestions} (P8), although \quote{a little bit of distraction is needed otherwise it's hard to think of creation in daily life} (P6). \\
\bottomrule
\end{tabular}
\end{table*}

\subsubsection{Factors Informing Future Real-World Viability: Perceived Benefits and Mechanisms}

\begin{figure}[h]
\centering
\includegraphics[width=1\textwidth]{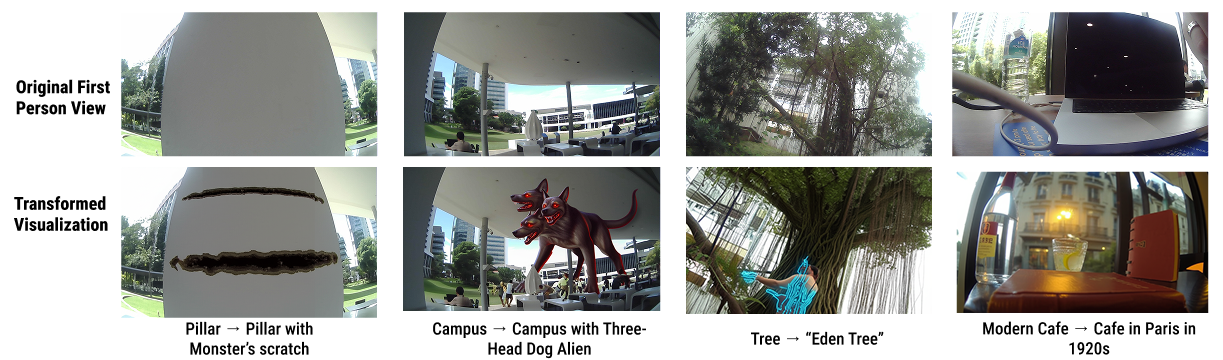}
\caption{Examples of transformed half-real/half-fiction visualizations in study 3.}
\label{fig:img_examples}
\end{figure}

We identified three factors that may inform the future viability of the \CRAFT{} approach, aligned with our three design goals (as shown in Table~\ref{tab:study3_unified_viability}):

\paragraph{(1) In-situ Inspiration Support (\DPone{}):} 
\label{sec:study3:results:rq1:dp1}
For \DPone{}, we investigated whether creating in-situ with \wearableAI{} could help writers generate novel, plot-relevant ideas from daily encounters. 
Participants consistently rated the probe highly for suggesting unexpected, useful ideas (M=6.0/7) and highlighting overlooked daily details that enriched their fiction (M=5.5/7). As shown in Table~\ref{tab:study3_unified_viability}-1, these benefits are driven by five mechanisms that form an ``inspiration lifecycle'': 1) AI \textit{Enhanced Noticing} of hidden sources; 2) \textit{Real-time Capture} of fleeting thoughts before they fade; 3) \textit{Serendipitous Plot Development} by reframing mundane events (e.g., P5's mosquito bite); 4) \textit{Contextual Knowledge Access} (e.g., linking plants to philosophy) to deepen their relevance; and 5) \textit{AR Visualization} to make abstract ideas tangible (Figure~\ref{fig:img_examples}). 
These observations suggest that \wearableAI{} can help turn real-world environments from ``places to write''~\cite{jabr2014walking} into active sources of fiction inspiration.

\paragraph{(2) Reality Utilization in Creative Work (\DPtwo{}):} 
\label{sec:study3:results:rq1:dp2}

For \DPtwo{}, we explored whether users could meaningfully integrate real-world content into their fiction, and our data (sec~\ref{sec:study3:results:overall_exp}) suggest all participants did so, with instances of such usage across various contexts and distributed micro moments in daily life (see Table~\ref{tab:study3_unified_viability}-2). Participants self-reported using personal daily observations on an average of 6.0/7, leading to a perceived \deepenEnvironment{} of 5.6/7. Notably, beyond single-element transformations, we observed \emph{holistic mapping}: writers systematically combined multiple environmental elements under coherent knowledge/theory structures to produce consistent story logic.

\paragraph{(3) Creative Enjoyment with Boundaries (\DPthree{}):} 
\label{sec:study3:results:rq1:dp3}

For \DPthree{}, participants reported meaningful creative enjoyment (M=5.8/7) and control over AI suggestions (M=5.2/7). As summarized in Table~\ref{tab:study3_unified_viability}-3, beyond the documentation efficiency from \DPone{}, participants derived enjoyment from two forms of \textit{immersive embodiment}: 1) \textit{first-person embodiment}, viewing the real world through their protagonist's eyes (P1), and 2) \textit{third-person dialogue}, role-playing with fictional characters (P7).
However, this immersion introduced two boundary concerns. First, potential \quote{emotional bleed}~\cite{hugaas2024bleed}—embodying characters in dark genres could affect real-world mood (P1)—advocates for moderation tools.
Second, despite initial boundary management in our probe (e.g., suggestion filtering based on user context), participants reported moderate distraction from primary tasks (\lowDistraction{}: M=3.9/7). While P6 noted that \quote{a little bit of distraction is needed} for daily creation, this score reflects a usability limitation driven by \quote{non-negligible AI interactions} and the \quote{prototype's [glasses frame] physical presence}. Future iterations should prioritize reducing cognitive load through more nuanced adaptive notifications and explicit focus mode toggles to better support the flow state essential for fiction writing.

\subsubsection{Diverse Creative Approaches in Real-world Use}
\label{sec:study3:results:rq2} 
\begin{figure}[htbp]
    \centering
    \includegraphics[width=1\textwidth]{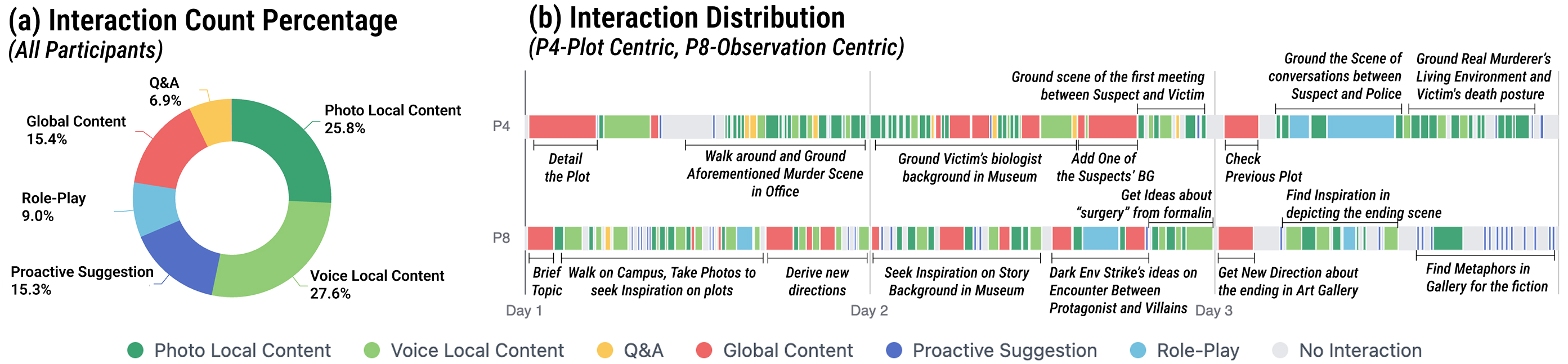}
    \caption{Usage Patterns of \CRAFT{} in Study 3: (a) interaction count percentage for all participants. Note: Global Content refers to global plot editing or checking, and Photo/Voice Local Content refers to individual fictional element authoring or updating. (b) interaction distribution (for P4 and P8) for each function with primary purposes marked up. All participants' distributions are in Appendix~\ref{appendix:study_data:real_world_testing} Figure~\ref{fig:all_distribution}.}
    \label{fig:study3:usage_patterns}
\end{figure}

Building on the factors above, we examined how writers applied these mechanisms in practice. 
We found that participants selectively leveraged these capabilities for different purposes, revealing two high-level creative workflows: structured plot-centric writing and exploratory observation-centric writing (Figure~\ref{fig:study3:usage_patterns}-b).

(1) \textit{Plot-centric creation} involves starting with pre-defined narrative structures and using the environment to ``fill in'' concrete details. As shown in Figure~\ref{fig:study3:usage_patterns}b (top), for her detective fiction, P4 first outlined the detailed murder plot (initial long red block on Day 1), then actively sought specific locations—such as office buildings and museum exhibits—to visualize and ``ground'' the characters' living environments, backgrounds, encounters, and death posture.

(2) \textit{Observation-centric creation} begins without fixed plots, allowing \wearableAI{}-augmented observations to trigger narrative direction. P8's timeline (Figure~\ref{fig:study3:usage_patterns}b, bottom) starts with a very brief topic expression (initial short red block in Day 1), followed by dense exploration on campus using photo capture (green blocks), leading to the formulation of a protagonist controlled by a \quote{knowledge chip.} On Day 2, her search for the chip's background in a natural history museum and serendipitous encounters with dark places inspired the conflict between the protagonist and the evil organization. Subsequently, observing specimens in formalin sparked the plot of a surgery to remove the \quote{knowledge chip,} while a Day 3 visit to an art museum inspired the protagonist's liberation and new life as an artist.

Crucially, writers can shift between styles. For instance, P4 incorporated serendipitous observations to refine narrative details (e.g., changing a suspect's occupation), while P8 transitioned to structured plotting once core ideas emerged (e.g., seeking details once the ending idea was confirmed at the beginning of Day 3). This adaptability suggests that sustained human--AI creative collaboration should support multiple writing workflows, rather than a single fixed process model.

\subsubsection{Challenges and Requirements for Sustained Creation}
\label{sec:study3:findings:additional_requirements}
While fully evaluating sustained viability requires longitudinal 
testing, Figure~\ref{fig:study3:quality_ratings} shows that author-perceived creation support
and interaction quality remained broadly stable or increased across three sessions, suggesting a positive potential for the sustained use of the \CRAFT{} approach. Nevertheless, the field trials revealed three key challenges for sustained creation that advocate further considerations for future systems:

\begin{figure}[ht]
    \centering
    \includegraphics[width=1\textwidth]{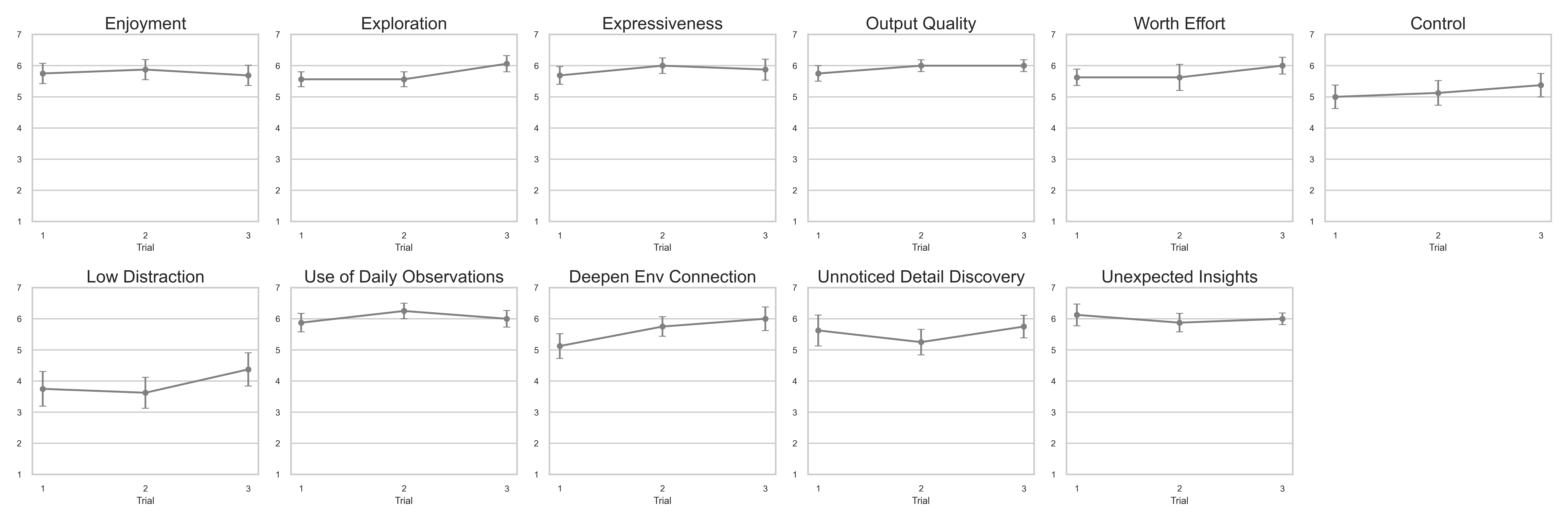}
    \caption{Quality ratings of the author-perceived creation support and interaction with the probe for all participants across three trial sessions.}
    \label{fig:study3:quality_ratings}
\end{figure}

\textbf{(1) Heterogeneous Creative Needs.} 
Writers' preferences varied by cognitive style and project stage, requiring \textit{High Configurability}. For \textit{suggestion style}, while P3 preferred \textit{symbolic} mappings (e.g., flowers representing emotions) for abstract ideation, P1 required \textit{identical} matches (e.g., a real café booth for environment description) to avoid unpredictability. Similarly, \textit{project lifecycles} dictated needs: early stages favored expansive suggestions, whereas P6 noted that later refinement required focused details (e.g., specific pastry aromas) to avoid derailing existing narratives.

\textbf{(2) Dual Fragmentation.} 
Unpredictable schedules led to two continuity issues, necessitating \textit{Dual Management Mechanisms}: 1) \textit{Sequence chaos} 
occurred when users captured snippets out of narrative order—for instance, P4 
drafted a climax in an office before developing the backstory in an art gallery. 
2) \textit{Context loss} occurred when writers left a location, causing memory 
fade that prevented them from recalling and continuing previous drafts upon 
revisiting.

\textbf{(3) Social Constraints.} 
Public usage exposed limitations in standard interaction, highlighting a need for \textit{Socially Compatible Interaction}. Voice input can sometimes be socially awkward—P2, for example, hesitated to dictate ``dystopian rebel plots'' in a quiet museum. Audio output could also pose risks of exposing sensitive content or disrupting social interactions among groups, underscoring the need for discreet, context-sensitive channels.

\section{OVERALL DISCUSSION}
\label{sec:discussion}

Through three complementary studies, we explored how the \CRAFT{} approach can support writers in transforming lived experiences into fictional narratives with \wearableAI{} that perceives, interprets, and creatively responds to real-world environments. Building on traditional practices of utilizing real-world materials, this approach highlights new opportunities for co-ideation with an AI \quote{writing companion} that senses the world alongside authors—facilitating serendipitous ideation, supporting authenticity in writing, and encouraging deeper real-world engagement. Our findings suggest three potential shifts in fiction writing with \wearableAI{}, which we discuss below alongside their broader implications for wearable creativity support.

\subsection{Potential Shifts in Fiction Writing Through the \CRAFT{} Approach}

\subsubsection{Content Shift: Reality-Grounding as a Potential Mitigation to Generic AI}
\label{sec:disc:content}

A commonly reported limitation of AI-assisted fiction writing emerged in our interviews: LLMs tend to generate generic content~\cite{kim2024authors, beguvs2024experimental, lee_empirical_2024} that lacks the authentic \quote{thickness} and \quote{lived feel} of fiction (P3, P7, sec~\ref{sec:study1:results:attitude:limitation}). This motivated \DPtwo{} (Promote authenticity triad), which uses real-world observations as \textbf{authenticating constraints} to ground AI outputs in specific, lived details. Our field trials suggest this grounding has the potential to counter genericness. With 215 real-world elements integrated across 8 stories, participants reported that their writing gained specificity and personal resonance when \wearableAI{} attended to concrete environmental details—such as blue-packaged cookies, broken pillars, or passersby's gestures. Notably, these elements carry embodied, personal meaning that AI alone cannot easily fabricate.

This points to a broader implication for human--AI creative collaboration: \textit{to counter generic AI content, systems could leverage reality as authenticating constraints that promote specificity and preserve personal voice}. Rather than relying solely on internal model knowledge, \wearableAI{} can be designed to interpret what is present and creatively transform it within an evolving fictional context. This shifts the AI's role from fabricating generic scenes toward \textit{translating} specific, lived moments, which may better preserve the \quote{human thickness} of creative output.

\subsubsection{Process Shift: From Disconnected Sessions to Distributed Micro-Creation In-situ}
\label{sec:disc:process}

Fiction writing typically separates inspiration (in-situ) from execution (ex-situ) due to high cognitive load and limited time~\cite{flower1981cognitive}. While \textit{micro-writing}~\cite{cai_pandalens_2024, teevan2016supporting} successfully leverages fragmented moments for documentation tasks like lifelogging~\cite{cai_pandalens_2024}, fiction writing requires active world construction and cross-session narrative consistency. Study~1 identified three barriers that limit the use of micro-moments for fiction: noticing narrative potential in everyday environments, deepening surface observations with fiction-relevant knowledge, and imagining transformations from raw observations. These barriers informed key design and probe mechanisms, including entity prioritization and perceptual augmentation (e.g., in proactive suggestion), real-time Q\&A grounded in fictional context, and half-real/half-fiction visualization for imagination scaffolding, enabling low-effort engagement through ring and voice interactions in daily life.

Our field trials suggest that lowering these barriers can extend \textit{micro-writing} into \textbf{micro-creation}: writers actively construct fictional worlds during brief moments rather than merely record inspiration. With each user-initiated interaction averaging 1.5 minutes (30.0 times/session) and proactive suggestions appearing for 0.5 minutes (5.4 times/session), participants identified hidden narrative cues, accessed context-specific knowledge, and imagined fictional scenes while walking or visiting museums. Across 24 sessions, they developed 8 complete stories, suggesting that distributed micro-creations could still accumulate into coherent character arcs, plot causality, and thematic unity that authors found satisfying. Participants also described this approach as helping overcome long-term time barriers (P1), reducing creative blocks through ``real-timeness'' (P3), and improving material collection and quality (P5).

We also observed different workflow preferences: experienced writers tended to use micro-creation mainly for ideation before later computer-based organization, whereas novice writers completed most tasks through micro-creation with minimal revision. This suggests that micro-creation can support multiple creative workflows across experience levels.

\subsubsection{Experiential and Ethical Shift: Additional Considerations for In-situ Creative Authoring}
\label{sec:disc:experience}

While reality-grounding and micro-creation open new possibilities, participants across all studies highlighted experiential and ethical trade-offs that concretize \DPthree{} (Preserve creative agency, enjoyment, and life-art boundaries).

\textit{Immersive Experiences and Psychological Safety.}
Writers valued immersive authoring features such as AR visualization and role-play for inhabiting character perspectives and deepening \quote{emotional truth}~\cite{jones2023embodying, bomba2024choreographer}. However, highly realistic immersion also introduced risks. In Study~2, P3 described an \quote{uncanny valley} effect~\cite{wang2015uncanny} when the AI superimposed a lifelike character onto her empty office, and in Study~3, P1 reported potential \textit{emotional bleed} when writing darker themes. These findings suggest that long-term systems should incorporate safeguards such as adaptive filtering of immersion intensity based on genre and mood~\cite{chen2022trauma, pataranutaporn2025synthetic}.

\textit{Navigating Social and Ethical Boundaries.}
Moving from private desks to public spaces shifts the ethical landscape of writing, introducing complex challenges regarding both \textit{how} writers interact (modality) and \textit{what} data they utilize (content).
First, regarding \textit{Interaction Modality}, writers sometimes hesitated to use voice input due to "social pressure and risks" (e.g., for sensitive/awkward plots in sec~\ref{sec:study3:findings:additional_requirements}). This necessitates \textit{Socially Compatible Interaction}: systems can support \textit{Discreet Input} (e.g., silent ring gestures for commonly used commands) and \textit{Context-Aware Output} (e.g., switching to visual text in crowds) to protect the writer's privacy.
Second, regarding \textit{Content and Data}, utilizing real-world data raises ethical concerns at multiple levels. The \textit{always-on} nature of context-aware wearables means continuous first-person capture may inadvertently surveil users and bystanders~\cite{denning2014situ, zhang2025through}, while reliance on cloud-based LLMs means sensitive creative content is transmitted externally. Since fiction writers often take real people for inspiration (sec~\ref{sec:study1_findings_principles}), future systems should adopt careful privacy design~\cite{zhang2025through, cai_pandalens_2024, cai_aiget_2025, rajaram_exploring_2025}, including \textit{Ethical Transformation}—automatically replacing identifiable attributes (faces, names) with AI-generated substitutes—as well as on-device processing, configurable capture zones, ambient bystander indicators, and encrypted pipelines to ensure data privacy. Furthermore, as content blends reality, human input, and AI generation, systems must ensure \textit{Provenance Clarity}~\cite{holopainen2025infinity, gero_social_2023}, explicitly highlighting grounded observations and algorithmic fabrications to preserve creative integrity.

\subsection{Additional Design Implications for Future Systems}

\subsubsection{Supporting Personalized and Adaptive Creation Needs}
All three studies revealed a common design requirement: \wearableAI{} needs to support multi-level personalization, adapting to a writer's long-term preferences, mid-term project stage, and situational needs. Specifically, the system must accommodate personalized material sources, writing style (sec~\ref{sec:study1_findings_principles}), and workflows (plot-centric vs.\ observation-centric, sec~\ref{sec:study3:results:rq2}) and adapt its support to the project's lifecycle (e.g., divergent suggestions in early stages and convergent, focused details during refinement, sec~\ref{sec:study3:findings:additional_requirements}). It must also offer granular controls on \textit{what} to suggest (e.g., similarity types), \textit{when} to intervene (proactive vs. on-demand vs. delayed), and \textit{how} to deliver feedback (support length \& depth) (sec~\ref{sec:study2_findings_how} \& \ref{sec:study3:findings:additional_requirements}) based on real-time needs. Ultimately, effective creative support requires not finding perfect default settings, but enabling writers to continuously tune their AI collaborator to match evolving needs.

\subsubsection{Bridging Fragmentation with Intelligent Continuity}

Our field trials suggest that fully realizing the potential of micro-creation in distributed sessions requires intelligent continuity support to address ``sequence chaos'' and ``context loss'' (sec~\ref{sec:study3:findings:additional_requirements}). Future systems should move beyond simple capture toward active contextual management. Two promising directions emerged: (1) \textit{Spatially Aware Memory}, leveraging geolocation to create ``location anchors'' that proactively resurface relevant drafts upon revisiting a place; and (2) \textit{Intelligent Sequence Organization}, whereby systems clarify vague structural commands (e.g., if a user asks to ``insert a scene before the library visit,'' the system must request clarification if multiple library scenes exist, such as, earlier ``book research visit'' vs. later ``finals study visit'') and synchronize seamlessly with desktop management tools~\cite{qin2025node} for post-capture refinement. These mechanisms would enable micro-creations to be effectively composed into coherent longer-form narratives.

\subsubsection{Using \CRAFT{} Design Mechanisms to Guide Future Implementations}

\begin{figure}[htbp]
    \centering
    \includegraphics[width=1\textwidth]{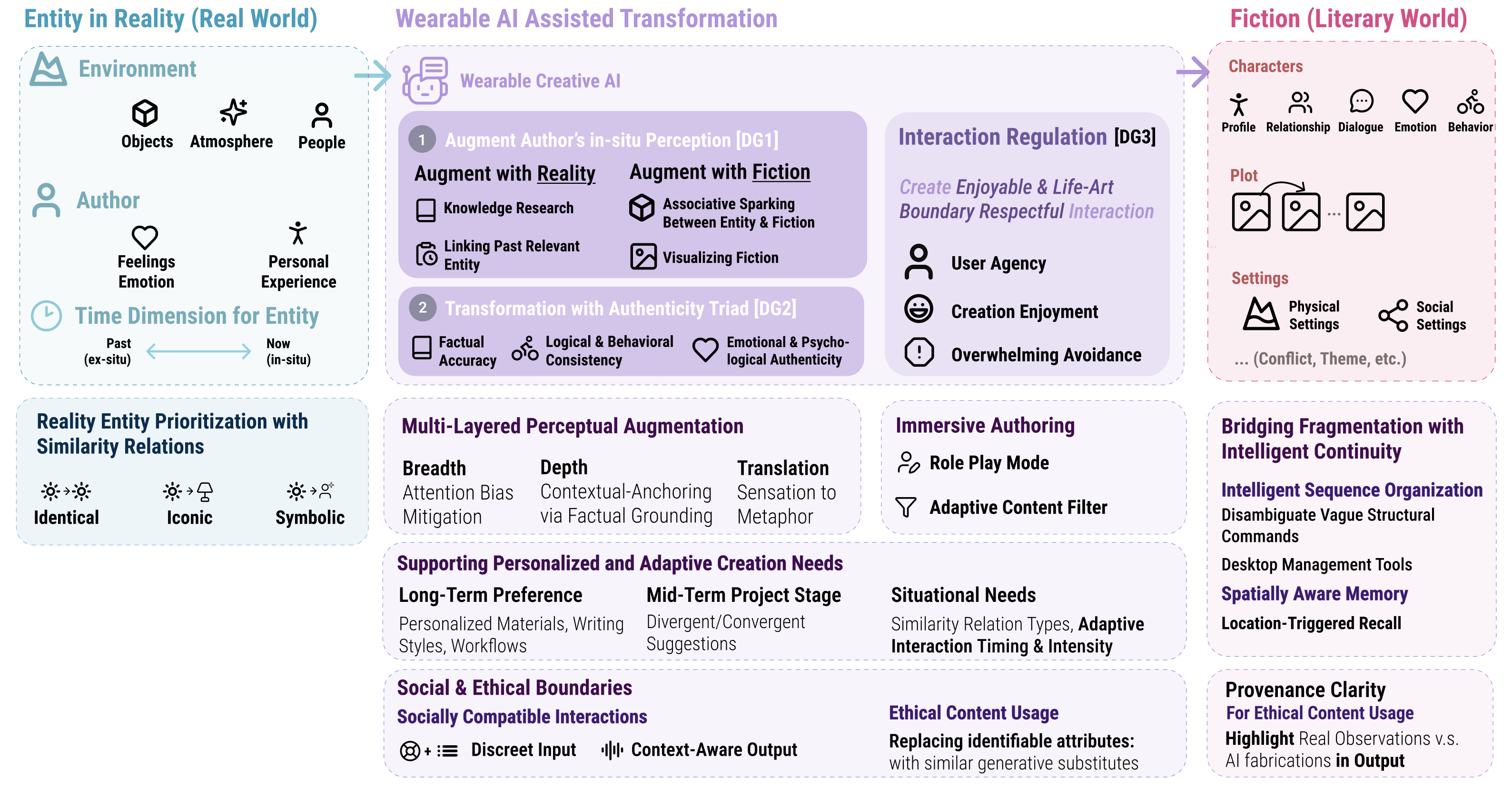}
    \caption{Design suggestion summary from all studies to guide future implementation.}
    \label{fig:design_summary}
\end{figure}

Figure~\ref{fig:design_summary} synthesizes the main design suggestions from three studies. It illustrates how \wearableAI{} can leverage real-world entities (environment and author context) and, guided by three design goals, transform them into fictional elements such as characters, plot, and settings. The figure highlights four key considerations for future implementations: (1) narrowing sensed input space using similarity relations (identical, iconic, symbolic); (2) supporting multi-layered perceptual augmentation and immersive authoring while preserving agency and avoiding overload; (3) maintaining continuity and provenance across fragmented sessions; and (4) respecting social, ethical, and personal boundaries during both input and output.

While derived from fiction writing, participants with cross-domain experience suggested that these mechanisms could extend to screenwriting, poetry, and interactive media. For instance, P2 (Study 2) noted that metaphor generation supports poetry, while P4 (Study 3) highlighted the value of in-situ role-play for drafting screenwriting dialogue in relevant environments. Future domain-specific adaptations might include visual storyboards for screenwriting, capturing rhythmic patterns of nature sounds for poetry, or olfactory feedback~\cite{de2022olfactory, gonccalves2017smell} for interactive media creation.

\section{LIMITATIONS}
\label{sec:limitations}

While \CRAFT{} offers an initial exploration of wearable creative AI for fiction writing, this work has several limitations.

\textit{Hardware and probe setup.} As a design exploration, this work implements a technology probe rather than a fully engineered system. Accordingly, reliability, energy consumption, and network robustness were not systematically optimized and evaluated in the present probe. Our probe also used a tethered setup rather than a lightweight standalone wearable. This hardware configuration could constrain spontaneity, comfort, and mobility in everyday use. In addition, speaking aloud in public introduced social barriers that may have limited engagement in quiet or socially sensitive settings. Future work should explore lower-friction input (e.g., whisper input~\cite{rekimoto2023wesper}) and more standalone wearable deployments. Our current MLLM pipeline also introduces moderate latency (\textasciitilde8s), which may make proactive suggestions feel interruptive in rapidly changing environments. Future systems could reduce this effect by using lighter-weight models or adaptive model selection in time-sensitive situations.

\textit{Study scope and evaluation.} This work is primarily an exploratory design study rather than an outcome-validation study. Study~3 was therefore relatively short and involved a small sample, partly due to the difficulty of securing multi-day participation from experienced writers; most participants also wrote primarily in English or Chinese. Longer-term deployments without experimenter presence, together with broader linguistic and cultural diversity, would better reveal sustained and more naturalistic use. Although we included descriptive text statistics, we did not conduct independent third-party evaluations of literary quality or baseline comparisons against alternatives (e.g., smartphone or desktop tools). To preserve a naturalistic creative process, we also did not require explicit accept/reject decisions for proactive suggestions; participants often synthesized multiple suggestions rather than adopting them verbatim, making post-hoc quantification of acceptance rates difficult. Future studies could use lightweight in-context queries to capture such data with minimal disruption and include baseline comparisons to better assess the relative value of wearable support over alternative tools. Given that creative writing judgments vary substantially across readers~\cite{marco2025reader}, larger and more diverse reader panels would further strengthen future evaluations.

\section{CONCLUSION}
\label{sec:conclusion}

We presented \CRAFT{}, an approach to using \wearableAI{} for in-situ fiction writing. Through three studies, we explored the desirability, feasibility, and potential viability of transforming lived experiences into fictional material. This process yielded three design goals: augmenting in-situ perception, promoting the authenticity triad, and preserving creative boundaries. We operationalized these goals into concrete interaction mechanisms, and supported field trials suggest that the \CRAFT{} approach could enable distributed micro-creation, allowing writers to bridge cognitive gaps through embodied authoring and transform serendipitous daily encounters and real-world details into author-satisfying narratives.
This work serves as an initial exploration, and realizing its full potential will require longitudinal deployments, comparative evaluations with external assessments of creative outcomes, and further design development to enable personalized, ethical, and sustainable usage.
As AR smart glasses and AI mature, \CRAFT{} extends the scope of wearable computing and outlines a practical path toward everyday, situated human–AI co-creation. This approach has the potential to make creative work more grounded in lived experience, enjoyable in practice, and accessible to a broader range of writers seeking to capture and transform the richness of daily life into compelling fiction.

\begin{acks}
This research was supported by the CityU Start-up Grant (9610677).
We thank the Lee Kong Chian Natural History Museum and the NUS Museum
for supporting our user studies, the participating writers for sharing
their experiences and feedback, and the reviewers for their constructive
comments.
\end{acks}


\appendix

\section{RECRUITMENT AND STOPPING RULE}
\label{appendix:recruitment}
For each study, sample size was determined during data collection using a
prespecified thematic-saturation stopping criterion. Recruitment ended when
two consecutive participant interviews or study sessions yielded no
substantively new themes.

\section{DATA ANALYSIS}
\label{appendix:study_analysis}
We analyzed the interview, workshop, and field-trial data using an inductive, reflexive thematic analysis adapted from Braun and Clarke’s guidelines~\cite{braun2006thematic, braun2021thematic}. To begin, two authors independently reviewed a subset of the data (three interview transcripts, two workshop sessions, and three supported field-trial sessions). Each author generated initial, data-driven codes, focusing on participants’ accounts and experiences. The two authors then met to discuss their interpretations and compare early codes to develop a shared analytic orientation. After this initial collaborative phase, one author coded the remaining data, iteratively updating and reorganizing the codes as new insights emerged. Themes were developed by the co-authors through multiple rounds of reviewing coded segments, grouping related ideas, and refining theme boundaries to ensure conceptual clarity and coherence. Finally, both authors revisited the full dataset—including transcripts, observational notes, and available video footage—to extract representative quotes and verify that the themes accurately reflected the breadth and depth of participants’ accounts.

\section{IMPLEMENTATION OF THE TECHNOLOGY PROBE}
\label{appendix:design_probe}

\subsection{Implementation Layer: Hardware and System Architecture}
The \CRAFT{} probe integrates Xreal Air smart glasses connected via USB-C to a MacBook Pro (M2 Pro), with a Pupil Core camera capturing first-person view (FPV) frames. The software stack combines a React/TypeScript frontend with WebSocket-based real-time communication to a FastAPI/Python backend. The backend orchestrates multiple AI services: Gemini 2.5 Flash for contextual text generation, Gemini 2.5 Flash-Lite/Flash for visual scene analysis, Gemini 2.0 Flash image generation model for image generation, Whisper WebGPU for offline speech transcription, and Gemini 2.5 Pro for summarizing fictional drafts. This modular architecture supports multimodal input (photo, speech, environmental sensing) across the probe pipeline. The MLLM pipeline latency reported in Section~\ref{sec:tech_prob:implementation} was measured during pilot testing over 20 proactive suggestion requests and 20 user-initiated query requests to characterize the current probe implementation.

\subsection{Interaction Layer: User and AI Initiatives}
The probe system implements a mixed-initiative interaction pipeline:
\begin{itemize}
  \item \textbf{User-initiative flows:} Triggered by explicit ring mouse interactions and voice commands (e.g., photo capture, voice query, role-play initiation). Responses are generated on demand and update the evolving fiction context. Post-session, users can utilize a desktop UI to generate plot node graphs and full drafts.
  \item \textbf{AI-initiative flows:} Triggered by continuous environmental monitoring (scene changes, contextual cues). These generate lightweight, proactive suggestions to stimulate creativity.
\end{itemize}
The smart glasses interface utilizes peripheral vision placement~\cite{cai2023paraglassmenu} to prevent interruption to primary tasks.

\subsection{Data and LLM Processing Layer: Context Management and Pipelines}

\subsubsection{Contextual Data Model}
The probe system maintains three synchronized contexts:
\begin{enumerate}
  \item \textbf{User preferences and goals} (creative objectives, style).
  \item \textbf{Environmental context} (visual scenes, location, time).
  \item \textbf{Fictional context} (structured models of plot, characters, settings, and style; plus latest full drafts), incrementally updated to ensure narrative consistency across sessions.
\end{enumerate}

\subsubsection{Proactive Suggestion Pipeline}
\label{appendix:proactive_pipeline}
The proactive suggestion pipeline consists of four key mechanisms:

\begin{itemize}
    \item \textbf{Triggering Logic:} The system operates on a fixed \textbf{30-second interval} loop. A new sensing cycle initiates every 30 seconds as the user moves through the environment; however, not every cycle results in an LLM request, as requesting is also controlled by the filtering mechanisms detailed below.
    \item \textbf{Information Flow:} The pipeline proceeds as follows: \textit{Sensing}  $\rightarrow$ \textit{Pre-filtering} $\rightarrow$ \textit{Context Fusion} $\rightarrow$ \textit{Generation} (Gemini 2.5 Flash) $\rightarrow$ \textit{Post-filtering} $\rightarrow$ \textit{Output}.
    \item \textbf{Context Fusion:} The prompt is constructed by integrating four data sources: (1) The current \textbf{Visual Stream} (the last 20 FPV frames covering 10 seconds) and geolocation; (2) The evolving \textbf{Fiction Context} (characters, setting, and style stored in JSON format); (3) The \textbf{Relevant Suggestion History} (top-10 historical suggestions whose associated environment description is similar to the latest environment) to ensure narrative continuity and prevent repetition; and (4) The \textbf{Full LLM Conversation History} of user-initiated actions within one creation session, utilized to infer the user's recent creative intent.
    \item \textbf{Failure \& Overload Handling:} To mitigate distraction and redundancy, the system employs a three-stage filter. First, a local \textbf{Visual Pre-filter} blocks MLLM requests when the current FPV remains visually similar to the context that triggered the last displayed suggestion, preventing unnecessary processing in static or visually repetitive environments. It uses both a frame-level MobileNet~\cite{howard2019searching} check, comparing the current frame with the saved anchor frame (cosine similarity $>0.8$), and a window-level check comparing the current FPV window with the saved FPV window (sequence similarity $>0.75$). Second, during generation, the MLLM performs an \textbf{Interruptibility Check}, returning \texttt{null} if the user's inferred cognitive load is high, using chain-of-thought style reasoning \cite{wei2022chain} and few-shot examples. Finally, a local \textbf{Post-filter} suppresses generated suggestions that are semantically similar (all-MiniLM-L6-v2~\cite{wang2020minilm}-based text embedding similarity $> 0.8$) to recent suggestion history.
\end{itemize}

\noindent\textbf{Prompt Structure:} The refined instruction (based on Study~2's insights) guides the generation process as follows:

{\footnotesize
\begin{verbatim}
# Role & Goal
You are a creative writing partner. Analyze the user's surroundings and ongoing story to offer proactive suggestions. 
# Core Instructions [From Original Full Prompt]
1. Priority: Quality over quantity. If no connection is natural, return null.
2. Analyze User State and Environment: Infer behaviors and key environmental elements.
3. Analyze Interaction History: Identify topics already suggested.
4. Similarity Priority Policy: Determine Identical (Indexical), Iconic, or Symbolic opportunities.
5. Identify Advanced Suggestion Techniques: Direct Reminder, Gap-Filling, and Metaphorical Enhancement.
6. Maintain Interaction Quality & Rhythm: Analyze user behavior and infer interruptibility. Return null if timing is inappropriate.
# Output Schema (JSON) {
  "user_behavior": "e.g., walking slowly, observing rain", 
  "environment_element": { "primary": [...], "peripheral": [...], "desc": "..." },
  "topics_already_covered": ["List of themes from history"],
  "rationale": "Reasoning for the suggestion, similarity type used, and interruptibility",
  "interruptibility": 0.0-1.0, 
  "suggestion": "Concise, plot-forward prompt (<25 words) or null",
  "similarity_type": "Identical/Iconic/Symbolic or null",
  "triggering_element": "The specific object/scene inspiring the prompt",
  "plot_link": "If applicable, which prior story event this connects to and why the current FPV matches."}
[Few-shot Examples...]
\end{verbatim}}

\subsubsection{User-Initiative Pipeline}
\label{appendix:main_query_pipeline}
The workflow for user-initiated interaction proceeds as follows:
\begin{itemize}
    \item \textbf{Triggering:} Initiated by a \textbf{Ring Mouse Interaction} paired with user speech (e.g., "What clue could be hidden here?").
    \item \textbf{Information Flow:} The system employs a \textbf{Hybrid Routing} mechanism. It first scans for explicit frontend signals (e.g., specific button clicks for "Role-Play"). Absent explicit signals, the system routes the voice transcription and captured photo to the MLLM. The model functions as a router, analyzing user intent to determine the interaction mode (e.g., \texttt{authoring} for scene transformation, \texttt{plot\_ideation} for brainstorming).
    \item \textbf{Context Fusion:} The MLLM prompt aggregates the writer's writing preference, \textbf{User  Query (Audio + Transcript)}, and a similar \textbf{Multimodal Context} used in the Proactive Suggestion pipeline (Visual Stream, Location, Time, Fiction Context, Interaction History). This ensures user queries remain grounded in both the narrative and physical environment.
    \item \textbf{Handling:} Mode-specific JSON schemas enforce structured outputs. For \texttt{authoring}, the system generates transformed scene descriptions and follow-up questions. For \texttt{plot\_ideation}, it returns plot summaries and elicitation questions. These outputs update the persistent fiction context using another LLM.
\end{itemize}

\noindent\textbf{Main System Prompt:} It demonstrates high-level main prompt structures for content generation:
{\footnotesize\begin{verbatim}
# Core Principles
- Brevity: Keep responses to 1-3 sentences per field unless the user requests more detail.
- Audio Priority: When audio and transcript conflict, prioritize audio as authoritative.
- User Intent: Always prioritize explicit user ideas, preferences, and requirements.
# Quality Requirements
1. Creative Adaptation: Respect the user's chosen fictional setting; adapt the real-world environmental features into the fiction.
2. Narrative Coherence: Continue logically from existing story content; ensure consistency; avoid plot holes.
3. Authenticity: Ensure natural dialogue and realistic character behaviors that are logically and emotionally authentic; 
   Ensure factual accuracy for professional/historical/technical content.
4. Grounded Details: Base narratives on specific user observations and sensory details.
The user's preferred writing style is: "[WRITING_STYLE]". History of user-created stories: "[STORY_HISTORY]"
# Mode Selection Logic
Analyze input to determine mode:
1. Authoring: Creating/transforming specific scenes or characters.
2. Plot Ideation: High-level brainstorming or global plot questions.
# Mode: Authoring
- User Intent Priority: Always prioritize explicit user ideas, requirements, or preferences.
- Check if the user asks direct factual questions first.
- If creating a new scene: 1) Summarize & Describe setting/mood. 2) Transform real entities -> fictional elements.
- If updating an existing scene: Only update new/changed content based on requests.
  - Output JSON: { "answer": "[if user asks questions]", "transformative_scene": "...", "plot": "...", 
    "transformative_characters": [...], "response_to_user": "..." }
# Mode: Plot Ideation
1) Generate a plot summary with the main theme/characters. 2) Ask elicitation questions to clarify user ideas.
- Output JSON: { "plot_summary": "...", "elicitation_questions": "..." }
[Few-shot Examples...]
\end{verbatim}}

\noindent\textbf{Role-Play Prompt:} The LLM manages role-play interactions via the following logic:

{\footnotesize\begin{verbatim}
You are an expert role-play AI. Analyze the user's input:
1. Determine the user's role and AI's role.
2. Extract and refine user's in-character dialogue (clean transcription errors, remove out-of-character phrases).
3. Generate an appropriate in-character AI response.
4. Create an image prompt for the AI character.
- Return JSON: {"user_role": "...", "user_dialogue": "Refined in-character speech", 
"ai_role": "...", "ai_dialogue": "...", "image_prompt": "..."}
\end{verbatim}}

\noindent\textbf{Image Generation Prompt Guidance:} To visualize the transformed fictional world, a generative pipeline processes the user's live FPV frame alongside the fiction context. For general narrative moments (i.e., for  \texttt{authoring} mode), the MLLM is instructed to preserve the real-world spatial layout while overlaying fictional elements (i.e., "half-real, half-fiction"). In Role-Play mode, the prompt instructs the MLLM to generate a 1:1 portrait of the AI character.

\subsubsection{Desktop Editing Pipeline}
Following in-situ sessions, the probe supports a desktop-based editing phase. This pipeline aggregates the accumulated fiction context (scenes, characters, plot points) and visualizes them as a plot node graph~\cite{qin2025node} to assist narrative structuring. When draft generation is triggered (via Gemini 2.5 Pro), the LLM synthesizes the edited plots into a cohesive narrative, adhering to the \textit{Authenticity Triad} principles (Factual, Logical, and Emotional \& Psychological).

\section{SUPPORTED FIELD TRIALS}
\label{appendix:study3_measure}

\begin{figure}[h]
    \centering
    \includegraphics[width=1\textwidth]{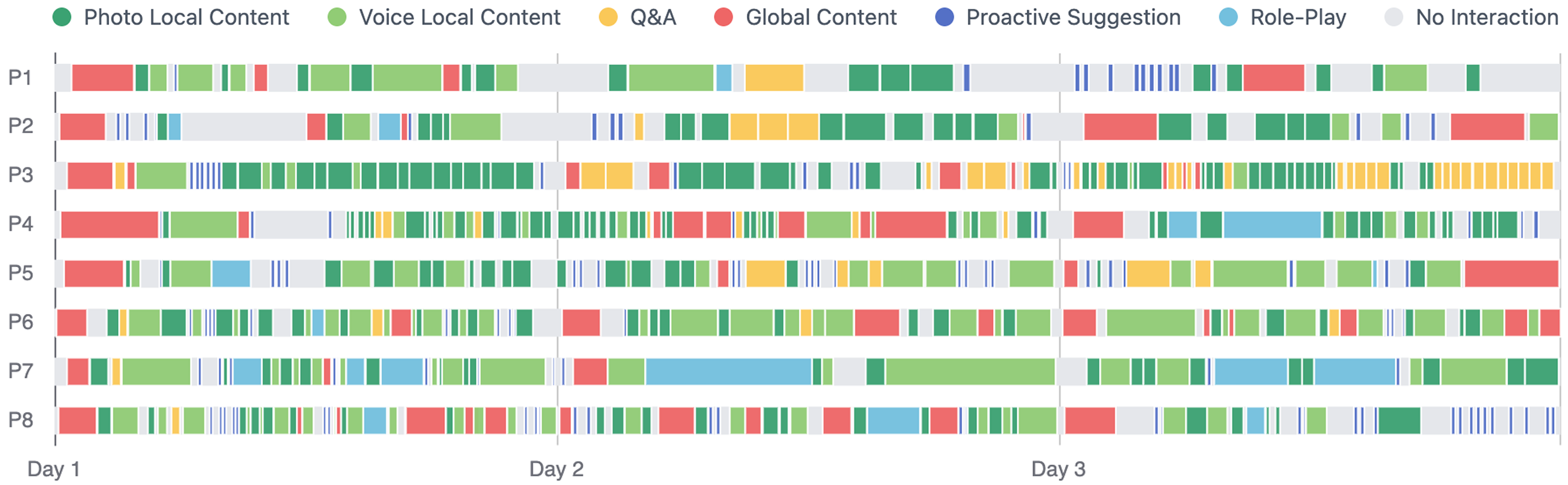}
    \caption{Interaction distribution across three sessions conducted on different days for all participants. A single colored block may represent multiple interactions if users refine the same content repeatedly.}
    \label{fig:all_distribution}
\end{figure}

Table~\ref{tab:study3_measures_ops} summarizes the subjective and objective measures used in supported field trials.
\label{appendix:study_data:real_world_testing}
Figure~\ref{fig:all_distribution} shows the distribution of interactions across three sessions conducted on different days for all participants in the supported field-trial study.

\begin{table*}
\footnotesize
\centering
\caption{Measures used in the supported field trial. Likert anchors: 1=Highly Disagree, 7=Highly Agree. Each CSI construct was computed as the within-session average of two items, except for Output Quality and Worth Effort, which were each assessed using a single item.}
\label{tab:study3_measures_ops}
\begin{tabular}{p{4.5cm}p{4.1cm}p{6.2cm}}
\toprule
\textbf{Category} & \textbf{Measure} & \textbf{Definition} \\
\midrule
\multirow{5}{4.5cm}{\textbf{Author-Perceived Creative Support \cite{cherry_csi_2014, carroll_csi_2009}}} 
& \exploration{} \,(1–7; 2 items; mean) & \emph{(i)} “The new writing approach was easy for me to explore many different options, ideas, or outcomes without a lot of tedious, repetitive interaction.” \newline \emph{(ii)} “The new writing approach was helpful in allowing me to track different ideas, outcomes, or possibilities.” \\
& \enjoyment{} \,(1–7; 2 items; mean) & \emph{(i)} “I enjoyed using the new interaction approach for writing.” \newline \emph{(ii)} “I would be happy to use such interactions on a regular basis.” \\
& \expressiveness{} \,(1–7; 2 items; mean) & \emph{(i)} “This new writing approach allowed me to be very expressive.” \newline \emph{(ii)} “I was able to be very creative while doing the activity with this new writing approach.” \\
& \outputQuality{} \,(1-7) & “I was satisfied with what I got out of the new writing approach.” \\
& \worthEffort{} \,(1–7) & “What I produced was worth the effort required to produce it.” \\
\midrule
\multirow{6}{4.5cm}{\textbf{In-Situ Interaction Experience \cite{cai_pandalens_2024, cai_aiget_2025, janaka2022paracentral}}} 
& \lowDistraction{} \,(1–7) & “I was \textbf{not} distracted too much by this tool during my daily activity.” \\
& \control{} \,(1–7) & “I feel I have control of the tool.” \\
& \unexpectedness{} \,(1–7) & “The system provides unexpected yet useful/interesting perspectives for my writing.” \\
& \unnoticedDetail{} \,(1–7) & “The system helps me identify previously unnoticed details in my daily life.” \\
& \deepenEnvironment{} \,(1–7) & “The system helps me build a deep connection with the environment.” \\
& \useOfDailyObservation{} \,(1–7) & “My personal experiences and daily observations were utilized in today’s creation.” \\
\midrule
\multirow{3}{4.5cm}{\textbf{Descriptive Session / Story Statistics}}
& \taskDuration{} \,(minutes) & Minutes of active in-situ creation in the session. \\
& Interaction logs & Counts and types of user--AI interactions during each session (e.g., fictional element management, global plot management, proactive suggestion, role-play dialogue, and Q\&A). \\
& Final narrative statistics & Words, scenes, characters, plot events, and instances of character speech or internal thought. \\
\bottomrule
\end{tabular}
\end{table*}

\end{document}